%% file: Alma_compendium_v8.tex
\documentclass[pdftex,usegraphicx,usenatbib]{mn2e}

\usepackage{graphicx}
\usepackage{natbib}
\usepackage{fixltx2e}
\usepackage{mathptmx}
\usepackage{amsmath}
\usepackage{amssymb}
\usepackage{wasysym}
\usepackage{url}
\usepackage{nohyperref}
\usepackage{times}
\usepackage{epsfig}
\usepackage{ulem}
\usepackage{deluxetable}

\input{defn_apj}

\begin{document}

\title[Massive molecular gas flows in central galaxies]{Driving massive molecular gas flows in central cluster galaxies with AGN feedback}\author[H.R. Russell et al.]
{\parbox[]{7.in}{H.~R. Russell$^{1,2}$\thanks{E-mail:
      helen.russell@nottingham.ac.uk}, B.~R. McNamara$^{3,4,5}$, A.~C. Fabian$^{1}$, P.~E.~J. Nulsen$^{6,7}$, F. Combes$^{8}$, A.~C. Edge$^{9}$, M. Madar$^{2}$, V. Olivares$^{8}$, P. Salom\'e$^{8}$, A.~N. Vantyghem$^{3,10}$\\ %F. Combes$^{6,7}$, A.~C. Edge$^{8}$, M.~T. Hogan$^{2,3}$, M. McDonald$^{9}$, P. Salom\'e$^6$, G. Tremblay$^{10,4}$, A.~N. Vantyghem$^2$ \\
    \footnotesize  %%%% Include Makun!  Probably people on A2052 proposal.
    $^1$ Institute of Astronomy, Madingley Road, Cambridge CB3 0HA\\
    $^2$ School of Physics \& Astronomy, University of Nottingham, University Park, Nottingham NG7 2RD\\
    $^3$ Department of Physics and Astronomy, University of Waterloo, Waterloo, ON N2L 3G1, Canada\\
    $^4$ Waterloo Centre for Astrophysics, Waterloo, ON N2L 3G1, Canada \\
    $^5$ Perimeter Institute for Theoretical Physics, Waterloo, Canada\\
    $^6$ Harvard-Smithsonian Center for Astrophysics, 60 Garden Street, Cambridge, MA 02138, USA\\
    $^7$ ICRAR, University of Western Australia, 35 Stirling Hwy, Crawley, WA 6009, Australia\\ 
    $^8$ LERMA, Observatoire de Paris, PSL Research University, College de France, CNRS, Sorbonne Univ., Paris, France\\
%    $^7$ College de France, 11 Pl. M. Berthelot, 75005 Paris\\
    $^9$ Department of Physics, Durham University, Durham DH1 3LE\\
    $^{10}$ Department of Physics \& Astronomy, University of Manitoba, Winnipeg, MB R3T 2N2, Canada\\
%    $^9$ Kavli Institute for Astrophysics and Space Research, Massachusetts Institute of Technology, 77 Massachusetts Avenue, Cambridge, MA 02139, USA\\
%    $^{10}$ Department of Physics and Yale Center for Astronomy \& Astrophysics, Yale University, 217 Prospect Street, New Haven, CT 06511, USA\\
  } }

\maketitle

% https://www.dropbox.com/s/1oi251iavv1wb9c/Alma_compendium_v3.pdf?dl=0
% mcnamara@uwaterloo.ca, acf@ast.cam.ac.uk, pnulsen@cfa.harvard.edu, alastair.edge@durham.ac.uk, ppymm2@nottingham.ac.uk, a2vantyg@uwaterloo.ca, francoise.combes@obspm.fr, Philippe.salome@obspm.fr, valeria.olivares@obspm.fr

\begin{abstract}
  \noindent We present an analysis of new and archival ALMA observations of molecular gas in twelve central cluster galaxies.  We examine emerging trends in molecular filament morphology and gas velocities to understand their origins.  Molecular gas masses in these systems span $10^9-10^{11}\Msun$, far more than most gas-rich galaxies.  ALMA images reveal a distribution of morphologies from filamentary to disk-dominated structures.   Circumnuclear disks on kiloparsec scales appear rare.  In most systems, half to nearly all of the molecular gas lies in filamentary structures with masses of a few $\times10^{8-10}\Msun$ that extend radially several to several tens of kpc.  In nearly all cases the molecular gas velocities lie far below stellar velocity dispersions, indicating youth, transience or both.  Filament bulk velocities lie far below the galaxy's escape and free-fall speeds indicating they are bound and being decelerated.  Most extended molecular filaments surround or lie beneath radio bubbles inflated by the central AGN.  Smooth velocity gradients found along the filaments are consistent with gas flowing along streamlines surrounding these bubbles. Evidence suggests most of the molecular clouds formed from low entropy X-ray gas that became thermally unstable and cooled when lifted by the buoyant bubbles.
  %This behaviour is consistent with molecular clouds forming when the ratio of the atmospheric cooling time% to its infall time approaches or falls below unity.
  Uplifted gas will stall and fall back to the galaxy in a circulating flow. The distribution in morphologies from filament to disk-dominated sources therefore implies slowly evolving molecular structures driven by the episodic activity of the AGN. 
\end{abstract}
%In one case (PKS\,0745), nuclear gas has been entirely removed or is absent, and instead lies in filaments whose masses exceed the molecular gas mass of the Milky Way.
%Multiple generations of radio bubbles, as observed in nearby clusters,
%would provide both the energy required to drive the massive gas flows
%and the formation timescale required for consistency with the limits
%on X-ray gas cooling.

\begin{keywords}
  galaxies:active --- galaxies: jets --- galaxies:evolution --- galaxies:clusters:intracluster medium
\end{keywords}

\section{Introduction}
\label{sec:intro}

Active Galactic Nuclei (AGN) release the energy of accreting material
as intense bursts of radiation or jets of relativistic
particles.  These energetic outbursts are observed to drive fast ($>500\kmps$)
outflows of ionized, neutral and molecular gas on pc to
kpc scales in the surrounding interstellar medium
(eg. \citealt{Morganti05,Morganti13,Nesvadba06,Nesvadba11,Feruglio10,Alatalo11,Rupke11,Sturm11}).
The gas flows may be driven by wide-angle winds launched from a luminous
accretion disk, via radiation pressure on dust or hot thermal winds,
or instead accelerated by radio jets, through direct collisions or
bubble buoyancy (for reviews see
\citealt{Veilleux05,Fabian12,King15}).  By driving cold gas from the
host galaxy, AGN restrict the fuel available for star formation and
engender the slowdown in massive galaxy growth since the peak of star
formation activity at $z=2-3$
(\citealt{Binney95,Silk98,DiMatteo05,Croton06,Bower06,Hopkins06}).  This
mechanism is also self-limiting, as the outflows also deprive the
supermassive black hole (SMBH) of fuel, and therefore is known as AGN
feedback.

Massive galaxies lie at the centres of nearby rich galaxy
clusters and represent a challenge for AGN feedback.  Without a heat source, the extensive hot gas atmospheres surrounding
these galaxies would cool rapidly and flood the galaxy with cold gas
and star formation (for reviews see
\citealt{Fabian94,PetersonFabian06}).  Encouragingly, the
central AGN in these galaxies are preferentially radio-loud and essentially ubiquitous
at the heart of cluster atmospheres with short radiative cooling times
(eg. \citealt{Burns90,DunnFabian06,Best07}).  Chandra X-ray
observations reveal large surface brightness depressions, typically
tens of kpc in size, where radio bubbles inflated by the jets have
carved out cavities in the hot atmosphere
(eg. \citealt{Boehringer93,Churazov00, McNamara00,FabianPer00}).  The jet power
can be estimated from the sum of the internal energy of the radio
bubbles and the work done displacing the hot X-ray gas.  For radio
bubbles dominated by relativistic particles, this is given by
$4PV/t_{\mathrm{age}}$, where $P$ is the ambient hot gas pressure, $V$
is the bubble volume and $t_{\mathrm{age}}$ is the rise time of the
bubble (eg. \citealt{Churazov00, Churazov02,McNamaraNulsen07}).  These bubbles
rise buoyantly through the intracluster medium and are still visible
as X-ray cavities even after the radio emission has spectrally aged
and faded at higher radio frequencies.  Studies of large cluster
samples have shown that this energy input by the radio bubbles
closely balances the radiative cooling losses from the surrounding hot
atmosphere (eg. \citealt{Birzan04,Rafferty06,DunnFabian06,Nulsen09}).
This observed balance, together with the prevalence of short central
radiative cooling times below a Gyr requires a feedback loop that
couples the AGN heating and gas cooling processes together (for a review see
\citealt{McNamara12}).

% Need to rethink the point here ...
% Key phase is cold gas?  Fuels star formation and black hole?
% Single dish observations reveal large molecular gas mass, preferentially detected in cool core clusters
% Obs. of nearby clusters with IRAM and ALMA Early science reveal filaments and tentative disks
% Now many more objects in this study?

A perfectly balanced feedback loop would fail.  Some gas must cool into molecular clouds and accrete onto
the galaxy and eventually onto the nuclear black hole to maintain it.
AGN activity, luminous cool gas nebulae and star
formation are preferentially detected in central cluster galaxies when
the radiative cooling time of the hot atmosphere falls below the
threshold value of $\sim 10^{9}\yr$
(\citealt{Rafferty08,Cavagnolo08}).  The cold molecular phase
likely dominates the cool gas mass in these systems.  Single dish sub-mm
observations of CO emission detected molecular gas masses in excess of
$10^{9}\Msun$ (\citealt{Edge01,Salome03}).  Like nebular emission, CO emission is detected preferentially
in systems whose atmospheric cooling times fall below $\sim 10^{9}\yr$ (\citealt{Pulido18}).
Furthermore, molecular gas mass is correlated with the X-ray gas mass measured on similar spatial scales within the galaxy (\citealt{Pulido18}). These
two relationships indicate that the bulk of the molecular gas cooled from the hot atmospheres.
However, the cooling time threshold only indicates a high likelihood of molecular gas being present.  It is not sensitive to the level or mass of molecular gas.  
Additional parameters are at play, which
are likely uplift behind the rising bubbles leading to thermally unstable cooling (\citealt{McNamara16}), and the mass of atmospheric gas available to cool (\citealt{Pulido18} and section \ref{sec:discussion} below).

However, the spatial structure of this cold gas could
only be resolved in a few of the nearest systems
(\citealt{Edge02,Salome04,Salome06}).  IRAM and SMA observations of
the massive reservoir of cold gas at the centre of the nearby Perseus
cluster revealed a filamentary structure
(\citealt{Salome06,Salome11,Lim08}).  The cold gas is spatially
coincident with extended filaments of soft X-ray, ionized and warm
molecular gas that are drawn up beneath radio bubbles
(\citealt{FabianFil03,HatchPer06,Salome06,Lim12}).

With the arrival of the ALMA observatory, the relationship between the
radio bubble activity and the cold gas reservoir can now be resolved
in detail in large samples of central cluster galaxies.  During the
Early Science phase, studies have necessarily focused on individual,
bright targets
(eg. \citealt{McNamara14,Russell14,Russell16,David14,Tremblay16,Vantyghem16}).
It was therefore initially difficult to draw many broader conclusions
given the variety and complexity of the detected structure or to
investigate potential correlations with jet power, X-ray gas mass and
cooling rates.  Here we present a uniform analysis of new and archival
ALMA observations for a dozen central cluster galaxies.  With this
larger sample, we identify the most prevalent morphological and
kinematical trends, investigate the origin of clear correlations with
radio bubble activity, the mechanism and the fate of the observed cold gas flows.

% Range of structure?  But individual targets.  Now some patterns emerging.  Trends in jet power.
% Number of issues, therefore we look at a large sample

We assume $H_0=70\kmpspMpc$, $\Omega_m=0.3$ and $\Omega_\Lambda=0.7$.
All errors are $1\sigma$ unless otherwise noted.

\section{Data reduction}

% Tables with observation details and redshift?
%\begin{deluxetable}{l c c c c c c c c}
\begin{table*}
\begin{minipage}{\textwidth}
\caption{Target and observation details.  References: [1] \citet{Russell14}, [2] \citet{McNamara14}, [3] \citet{David14}, [4] \citet{Russell16}, [5] \citet{Tremblay16}, [6] \citet{Vantyghem16}, [7] \citet{Russell17}, [8] \citet{Vantyghem17}, [9] \citet{Russell17A1795}, [10] \citet{Vantyghem18}, [11] \citet{Tremblay18}.}
\begin{center}
\begin{tabular}{l c c c c c c c c}
\hline
%\tablecolumns{9}
%\tablewidth{0pc}
%\tablecaption{Target and observation details.}
%\tablehead{
%\colhead{Target} & \colhead{Redshift} & \colhead{Scale} & \colhead{CO line} & \colhead{$\nu_{\rm obs}$} & \colhead{On source} & \colhead{Obs. date} & \colhead{No. of} & References \\
%  & & \colhead{(kpc/$\arcsec$)} & & \colhead{(GHz)} & \colhead{(min)} & & \colhead{Antennas} & }
%  \startdata
Target & Redshift & Scale & CO line & $\nu_{\rm obs}$ & On source & Obs. date & No. of & References \\
& & (kpc/$\arcsec$) & & (GHz) & (min) & & Antennas & \\
  \hline
  A2052 & 0.0345 & 0.7 & J=2-1 & 222.856 & 82.0 & 2016-08-11, 2016-08-23 & 37, 35 & \\
%  & & & & & & 2016-08-23 & 35 \\
  PKS0745-191 & 0.1028 & 1.9 & J=1-0 & 104.526 & 25.3 & 2014-04-26, 2014-04-27 & 34, 36 & [4]\\
   & & & J=3-2 & 313.562 & 56.4 & 2014-08-19 & 32 & [4] \\ 
  A1795 & 0.0633 & 1.2 & J=2-1 & 216.822 & 72.0 & 2016-06-11, 2016-06-14 & 38 & [9] \\
  A2597 & 0.0821 & 1.5 & J=2-1 & 213.047 & 189.9 & 2013-11-17, 2013-11-18 & 29, 28 & [5] \\
  & & & & & & 2013-11-19 & 28 & \\
  NGC5044 & 0.0093 & 0.2 & J=2-1 & 228.440 & 23.3 & 2012-01-13 & 18 & [3] \\
  2A0335+096 & 0.0346 & 0.7 & J=1-0 & 111.416 & 35.9 & 2014-07-22, 2015-03-08 & 33, 30 & [6]\\
         & & & J=3-2 & 334.232 & 34.3 & 2014-08-12 & 34 & [6]\\
  RXJ0821.0+0752 & 0.111 & 2.0 & J=1-0 & 103.754 & 87.0 & 2016-10-30, 2016-11-04 & 41, 43 & [8] \\ % AVantyghem has z=0.109 but nu suggests 0.110 and cites Crawford'95 who have z=0.110 ... probably better to be consistent with Adrian?  He has line centre close to 0km/s (rather than at 270km/s).  Difference in redshift is exactly 300km/s so shift plots?
        & & & J=3-2 & 311.248 & 22.7 & 2016-10-01 & 41 & [8] \\
  RXCJ1504.1-0248 & 0.2169 & 3.5 & J=1-0 & 94.725 & 154.0 & 2016-10-27, 2016-10-27 & 40, 41 & [10]\\
  & & & & & & 2016-11-02, 2017-05-10 & 38, 47 & \\
        & & & J=3-2 & 284.161 & 34.5 & 2017-07-04 & 45 & [10] \\
  Phoenix & 0.596 & 6.8 & J=3-2 & 216.664 & 58.5 & 2014-06-15, 2014-06-16 & 34, 35 & [7] \\
  A1664 & 0.128 & 2.3 & J=1-0 & 102.191 & 50.4 & 2012-03-27, 2012-04-07& 15, 17 & [1] \\
  & & & J=3-2 & 306.557 & 70.6 & 2012-03-28, 2012-03-28 & 17, 16 & [1] \\ % Above theoretical rms (probably due to lots of flagging ..)
  A1835 & 0.252 & 3.9 & J=1-0 & 92.070 & 59.5 & 2012-03-27, 2012-04-07 & 15, 17 & [2] \\
  & & & J=3-2 & 276.190 & 59.5 & 2012-03-28, 2012-04-24 & 16, 20 & [2] \\
  A262 & 0.0162 & 0.33 & J=2-1 & 226.863 & 11.1 & 2016-06-27 & 42 & \\ % Can't find stellar absorption measurement.
%  \enddata
%  \tablecomments{References: [1] \citet{Russell14}, [2] \citet{McNamara14}, [3] \citet{David14}, [4] \citet{Russell16}, [5] \citet{Tremblay16}, [6] \citet{Vantyghem16}, [7] \citet{Russell17}, [8] \citet{Vantyghem17}, [9] \citet{Russell17A1795}, [10] \citet{Vantyghem18}, [11] \citet{Tremblay18}}
\hline
\end{tabular}
\label{tab:det}
\end{center}
\end{minipage}
\end{table*}
%  \label{tab:det}
%\end{deluxetable}
%\clearpage
%\newpage
%\begin{deluxetable}{l c c c c c c}
\begin{table*}
\begin{minipage}{\textwidth}
\caption{ALMA cube specifications for each target.}
\begin{center}
\begin{tabular}{l c c c c c c}
\hline
%\tablecolumns{7}
%\tablewidth{0pc}
%\tablecaption{ALMA cube specifications for each target.}
%\tablehead{
%%%%\colhead{Target} & \colhead{CO line} & \colhead{$\nu_{\rm obs}$} &\colhead{Beam} & \colhead{PA} & \colhead{Binning} & \colhead{rms} \\
%%%% & & \colhead{(GHz)} & \colhead{($\arcsec\times\arcsec$)} & \colhead{(deg)} & \colhead{($\kmps$)} & \colhead{($\mJypbm$)}}
Target & CO line & $\nu_{\rm obs}$ & Beam & PA & Binning & rms \\
& & (GHz) & ($\arcsec\times\arcsec$, $\mathrm{kpc}\times\mathrm{kpc}$) & (deg) & ($\kmps$) & ($\mJypbm$) \\
\hline
% \startdata
  A2052 & J=2-1 & 222.856 & $2.2\times1.2$, $1.5\times0.8$ & $89.9$ & 15 & 0.8 \\ 
  PKS0745-191 & J=1-0 & 104.526 & $1.6\times1.2$, $3.0\times2.3$ & $79.6$ & 10 & 0.6 \\
   & J=3-2 & 313.562 & $0.3\times0.2$, $0.6\times0.4$ & $78.3$ & 10 & 1 \\ 
  A1795 & J=2-1 & 216.822 & $0.8\times0.6$, $1.0\times0.7$ & $-15.3$ & 10 & 0.64 \\
  A2597 & J=2-1 & 213.047 & $1.0\times0.8$, $1.5\times1.2$ & $89.8$ & 10 & 0.4 \\
  NGC5044 & J=2-1 & 228.440 & $2.2\times1.4$, $0.4\times0.3$ & $-31.1$ & 10 & 1.5 \\
  2A0335+096 & J=1-0 & 111.416 & $1.1\times0.9$, $0.8\times0.6$ & $-34.9$ & 20 & 0.5 \\
         & J=3-2 & 334.232 & $0.4\times0.2$, $0.3\times0.1$ & $-50.4$ & 10 & 0.8 \\
  RXJ0821.0+0752 & J=1-0 & 103.754 & $0.7\times0.7$, $1.4\times1.4$ & $-31.4$ & 5 & 0.5 \\
        & J=3-2 & 311.248 & $0.2\times0.1$, $0.4\times0.2$ & $36.9$ & 5 & 1.3 \\
  RXCJ1504.1-0248 & J=1-0 & 94.725 & $0.9\times0.8$, $3.2\times2.8$ & $-82.2$ & 20 & 0.2 \\
        & J=3-2 & 284.161 & $0.3\times0.2$, $1.1\times0.7$ & $-57.0$ & 10 & 0.6 \\
  Phoenix & J=3-2 & 216.664 & $0.6\times0.6$, $4.1\times4.1$ & -37.9 & 12 & 0.3 \\
  A1664 & J=1-0 & 102.191 & $1.6\times1.3$, $3.7\times3.0$ & -89.9 & 40 & 0.5 \\
  & J=3-2 & 306.557 & $0.6\times0.4$, $1.4\times0.9$ & -83.7 & 30 & 1.5 \\ % Above theoretical rms (probably due to lots of flagging ..)
  A1835 & J=1-0 & 92.070 & $1.7\times1.3$, $6.6\times5.1$ & -83.3 & 5 & 1.1 \\
  & J=3-2 & 276.190 & $0.6\times0.4$, $2.3\times1.6$ & -80.0 & 5 & 3.0 \\
  A262 & J=2-1 & 226.863 & $1.0\times0.6$, $0.3\times0.2$ & 10.9 & 5 & 1.4 \\ % Can't find stellar absorption measurement.
%  \enddata
%  \tablecomments{}
\hline
\end{tabular}
\label{tab:cube}
\end{center}
\end{minipage}
\end{table*}
%\end{deluxetable}
%\clearpage
%\newpage

% Sample selection
We analysed new and archival ALMA observations of CO line emission in
central cluster galaxies to investigate emerging trends in the
molecular gas structure and kinematics.  Targets were principally
selected from single object studies in the literature (see Table
\ref{tab:det}) with the addition of new ALMA observations of the
central galaxy in Abell 2052 and archival observations of NGC\,708 at
the centre of A262.  This produced a total sample of 12 central
cluster galaxies.  The sample spans a wide range in molecular gas mass
($10^{7}-10^{11}\Msun$), X-ray cavity power ($10^{42}-10^{46}\ergps$),
star formation rate (a few to $\sim500\Msunpyr$) and redshift (up to
$0.596$).  However, predominantly bright, gas-rich systems were
preferentially, and necessarily, selected as early ALMA targets.  They
are therefore over-represented in our sample.  This sample is
neither complete nor unbiased and we are careful to consider
this throughout our analysis.

For each target, we selected CO line observations from the ALMA
archive as detailed in Table \ref{tab:det}.  The central galaxies were
observed with the ALMA $12\m$ array at frequencies corresponding to
the CO(1-0), CO(2-1) or CO(3-2) rotational transition lines with
additional spectral windows used to image the sub-mm continuum
emission.  Each dataset consists of a single pointing centred on the
galaxy nucleus and covers the ionized gas peak and the brightest
filaments.  The ALMA field of view ranges between $\sim1\amin$ in band
3 at $\sim100\GHz$ and $\sim20\asec$ in band 7 at $\sim300\GHz$.  The
spatial resolution and sensitivity vary significantly from the
earliest Cycle 0 observations (eg. A1664 and A1835) with $\sim15$
antennas and baselines up to $400\m$ to the latest datasets
(eg. RXCJ1504.1-0248) with $>40$ antennas and baselines up to a few
km.  This variation is reflected in the synthesized beam size and rms
of the final data cube as detailed in Table \ref{tab:cube}.

% We also include new ALMA observations of the central galaxy in Abell
% 2052 (UGC9799), which were taken in band 6 on 11 and 23 August 2016
% for a total time on source of $82\min$.  The single pointing was
% centred on the nucleus and covered the H$\alpha$ emission peak and
% inner ionized gas filaments beneath the N radio bubble
% (eg. \citealt{McDonald10,Blanton11,Balmaverde18}).  One spectral
% window covered the CO(2-1) emission line at $222.856\GHz$ and three
% additional windows at $220.97$, $235.96$ and $237.98\GHz$ were used to
% sample the sub-mm continuum emission.  Thirty-seven and thirty-five
% $12\m$ antennas were available for each observation respectively and
% baselines ranged from $15-1500\m$.  J1550+0527, J1511+0518 and
% J1337-1257 were observed for bandpass and flux calibration and
% observations of the phase calibrator J1505+0326 were interspersed with
% the target observations.  Frequency division correlator mode with
% bandwidth $1.875\GHz$ and resolution $977\kHz$ ($\sim1.2\kmps$) was
% used for the spectral window covering the CO(2-1) line.  However, the
% velocity channels were binned up to increase signal-to-noise with
% $15\kmps$ resolution for the final data cube (Table \ref{tab:cube}).
% The continuum spectral windows had much broader intrinsic spectral
% resolution of $15.6\MHz$ ($\sim20\kmps$).

\begin{figure*}
\begin{minipage}{\textwidth}
\centering
\includegraphics[width=0.98\columnwidth]{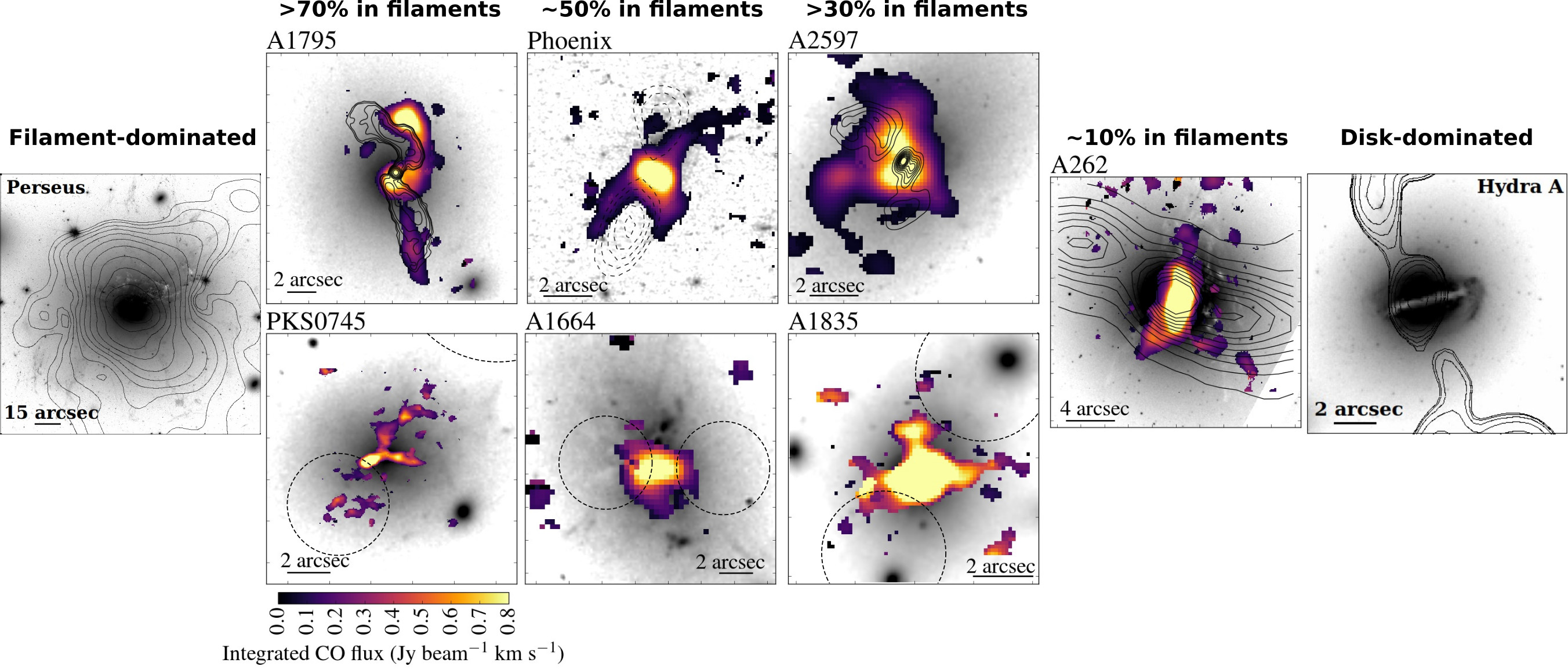}  
\caption{The molecular gas morphology spans a range from filament-dominated sources, for which the archetype is the Perseus cluster, to disk-dominated sources like Hydra A.  This is demonstrated by the selected subset of our sample.  Optical images from HST of the central galaxy's stellar light are shown in greyscale.  The integrated CO flux is shown in colour where detected at $>3\sigma$ and the same colour bar (lower left) applies to each image.  VLA radio contours are shown by the solid black lines.  X-ray surface brightness depressions corresponding to cavities are shown by the dashed black contours or schematically by the dashed black circles.}
\label{fig:montageflux}
\end{minipage}
\end{figure*}
\begin{figure*}
\begin{minipage}{\textwidth}
\centering
\includegraphics[width=0.98\columnwidth]{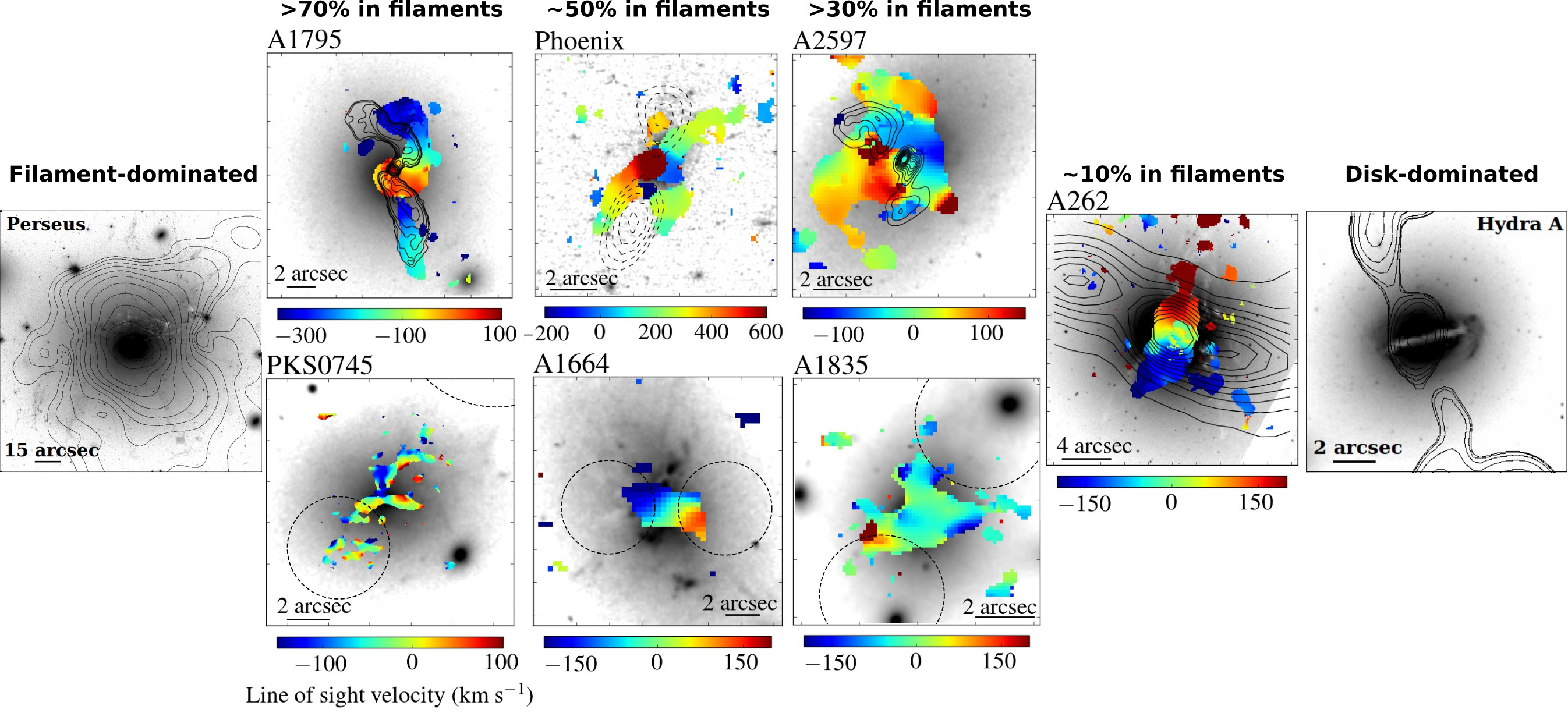}  
\caption{The molecular gas exhibits smooth velocity gradients along the filaments drawn up around or beneath radio bubbles or across circumnuclear gas disks.  Optical images from HST of the central galaxy's stellar light are shown in greyscale.  The line of sight velocity for the dominant component of the CO emission is shown in colour, where it is detected at $>3\sigma$.  VLA radio contours are shown by the solid black lines.  X-ray cavities are shown by the dashed black contours or schematically by the dashed black circles.}
\label{fig:montagevel}
\end{minipage}
\end{figure*}

% Data reduction
Each ALMA dataset was calibrated using the required version of
\textsc{casa} (\citealt{McMullin07}), which ranged from version 3.3.0
for the Early Science data to version 4.7.0 for the latest datasets.
Datasets taken in the early ALMA cycles were manually calibrated with
tailored data reduction scripts generated by ALMA staff while later
datasets were reduced by the automated ALMA science pipeline.  Several
datasets did require additional flagging and other modifications, for
example to the phase centre (eg. A1835) and total flux calibration
(eg. Phoenix) and to limit the impact of poor phase solutions
(eg. 2A0335+096, \citealt{Vantyghem16}).

Continuum-subtracted data cubes were generated using the \textsc{casa}
tasks \textsc{uvcontsub} and \textsc{clean}.  Different weightings
were tested for each dataset to determine the optimum for imaging.
Briggs weighting with a robust parameter of 2 (close to natural weighting) was favoured for targets
with extended filaments to provide the highest signal-to-noise in
these structures.  Otherwise, Briggs weighting with a robust parameter
of 0 was used to produce a good compromise between spatial resolution
and sensitivity.  We note that ALMA's good uv coverage ensured that
images generated with a range of weightings did not show any major
differences.  The rms in each final datacube was compared with and
found to be close to the corresponding theoretical rms, which is
dependent on the array configuration, integration time, frequency and
atmospheric conditions.  The synthesized beam size, velocity binning
and rms in each final datacube are detailed in Table \ref{tab:cube}.

Continuum images were produced from line-free channels in each
baseband and using uniform weighting.  Self-calibration was also employed
to produce modest increases in continuum sensitivity for targets
with continuum peaks greater than a few mJy.  The continuum emission
was due to nuclear point sources coincident with the AGN in all
targets, except RXJ0821+0752 where the sub-mm continuum is spatially
extended and offset from the AGN (\citealt{Vantyghem19}).

% velocity centre?

\section{Results}

Maps of the integrated intensity, velocity centre and FWHM were
produced for all twelve targets in our sample covering each CO line
observed with ALMA.  Integrated intensity maps were generated by
integrating over the CO line profile in each spatial pixel (zeroth
moment map).  A subset of the integrated intensity maps are shown in
Fig. \ref{fig:montageflux} overlaid on optical images of the central
galaxy's stellar light, with radio contours, X-ray contours or regions
showing the position of the X-ray cavities or radio bubbles.  All CO images for each
source are shown separately in the appendix together with optical and
X-ray images.

Maps of the gas velocities and line widths were generated by
extracting spectra from each cube in synthesized beam-sized regions
centred on each spatial pixel across the field of view.  Each
extracted spectrum was fitted with a model consisting of one, two and
then three Gaussian components using \textsc{mpfit}.  At least
$3\sigma$ significance was required for the detection of an emission
line in each region analysed.  The significance was assessed by
evaluating the number of false detections in Monte Carlo simulations
based on a null hypothesis model with no emission line and the
observed rms.  Fig. \ref{fig:montagevel} shows the velocity structure
of the dominant component for a subset of the sample and maps of the
best-fit line centre, FWHM and integrated intensity are shown for the
full sample in the appendix (Figs. \ref{fig:a2052} to
\ref{fig:A262maps}).  We also extracted spectra from regions covering
the full extent of the emission and individual structures, such as
filaments.  These spectra were similarly fitted with multiple Gaussian
components as required and the best-fit parameters are detailed in
the appendix.

In several systems, the integrated intensity maps show distinct
filaments of molecular gas that can be simply separated spatially.
Note that the term `filament' is used to describe an elongated
molecular structure, at least a few kpc in length, that extends
approximately radially from the galaxy centre (eg. PKS\,0745, Phoenix,
A1795).  These filaments are not resolved in our observations and
likely consist of many smaller strands (eg. Perseus cluster,
\citealt{Fabian08,GendronMarsolais18}).  We also identify clear and
possible kpc-scale circumnuclear molecular gas disks in several
systems where the emisson peak of a kpc-scale molecular structure is
centred on the nucleus and the velocity structure is consistent with
rotation about the galaxy's systemic velocity.  Whilst nearby Hydra A
clearly hosts an edge-on molecular gas disk (\citealt{Rose19}), the
molecular emission peaks at the centre of the majority of our more
distant targets are not resolved (eg. A1835, Phoenix).  Higher
resolution ALMA observations may therefore reveal smaller scale
circumnuclear gas disks.

% Can also separate spectrally
% Determine mass for each structure by separating spatially or spectrally and then calculate filament fraction

Molecular structures that are superimposed on the sky can instead
often be separated spectrally.  For example in A1664 and Phoenix, an
additional velocity component at the galaxy centre is consistent with
a circumnuclear gas disk.  In A2597, although the majority of the
molecular structure is difficult to disentangle, a second velocity
component is detected in the emission around the NE and S radio lobes,
which could indicate entrained gas.  The
CO line flux is detailed in the appendix for each molecular structure that could be cleanly
separated spatially and/or spectrally for each target and the
unresolved emission centred on the nucleus.

\subsection{Molecular gas morphology}
\label{sec:morphology}

% Need to classify structure into groups (eg. exclusively around cavities, behind cavities, some emission round cavities, ambiguous structure)
% Fraction of gas in filaments (caveat on resolution and optical thickness), typical extent (with caveats on max. scale and sensitivity), coincident with soft X-ray and H-alpha (star formation?)
% Patchy shell idea
% (Central peaks show limited evidence of rotation and disks => velocity)
% Offsets in peak from nucleus

The molecular gas morphology ranges from filament-dominated sources,
for which the archetype is the Perseus cluster
(eg. \citealt{Salome06,Salome11}), to disk-dominated sources, such as
Hydra A (eg. \citealt{Hamer14}, \citealt{Rose19}).  The ALMA targets
form a continuous distribution within these extremes
(Fig. \ref{fig:montageflux}).  In A1795 and PKS\,0745, $>70\%$ of the
molecular gas reservoir lies in extended filaments.  In Phoenix and
A1664, the molecular gas is split more evenly between a possible
circumnuclear disk (see section \ref{sec:vel}), and
extended filaments.  For A2597 and A1835, the filaments and
circumnuclear structures are more difficult to cleanly disentangle and
we estimate that at least a third of the molecular gas lies in the
filaments.  By comparison, in A262 and Hydra A, the majority of the
molecular gas is resolved in a rotating circumnuclear gas disk
(eg. \citealt{Prandoni07,Hatch07}) and only $\sim10\%$ extends to
larger radii.  Disk-dominated sources are rare in our sample
($\sim15\%$) with only these two clear examples.  Improved
spatial resolution may increase estimates of the fraction of molecular
gas in the filaments (eg. A1835 and A2597) and could reveal smaller scale, low mass circumnuclear gas disks (eg. Perseus, \citealt{Scharwachter13,Nagai19}).

% Note that the extended fraction will likely be a lower limit as the limited
% spatial resolution in several cases will produce an underestimate.
% For example, the Early Science observations of A1835 do not resolve
% the majority of the molecular gas, which lies in the central peak.

Filaments are entrained around or clearly extend toward X-ray
cavities or radio bubbles in at least six of the twelve targets (A1795, Phoenix,
PKS\,0745, A1835, A2597, A262).  In PKS\,0745, for example, the SE
filament extends toward the SE cavity, the N filament extends towards
the more distant NW cavity.  The SW filament is aligned with a lobe of
radio emission that may originate in a new AGN outburst
(\citealt{Russell16}).  The extended molecular emission coincides
exclusively with the soft X-ray rim of a cavity and ionized gas
filaments drawn up around this structure in a further three objects
(RXJ\,0821, RXCJ\,1504, A2052).  Whilst molecular structures in the
remaining targets, 2A\,0335, A1664 and NGC\,5044, are apparently
elongated toward X-ray cavities, the mismatch in spatial scales probed
by ALMA and \textit{Chandra} is too large to conclude a clear link
(\citealt{David14,Vantyghem16,Calzadilla18}).

\begin{figure*}
\begin{minipage}{\textwidth}
  \centering
\raisebox{0.2cm}{\includegraphics[width=0.3\columnwidth]{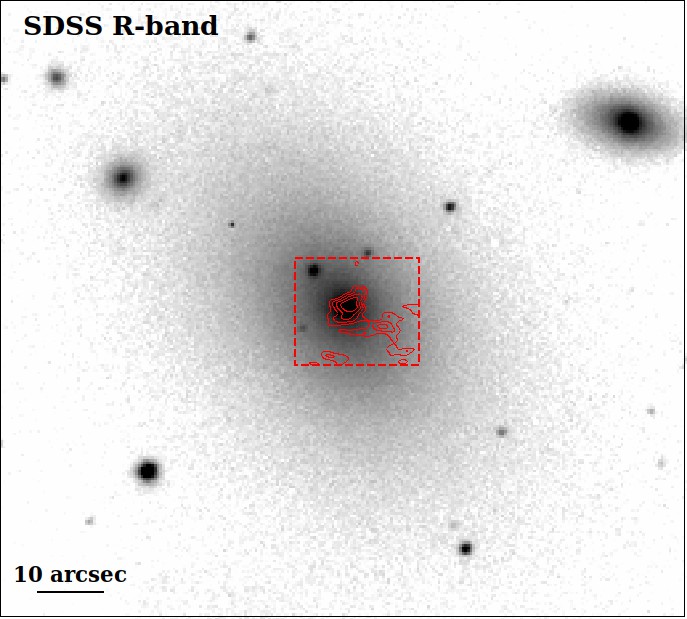}}
\raisebox{0.2cm}{\includegraphics[width=0.3\columnwidth]{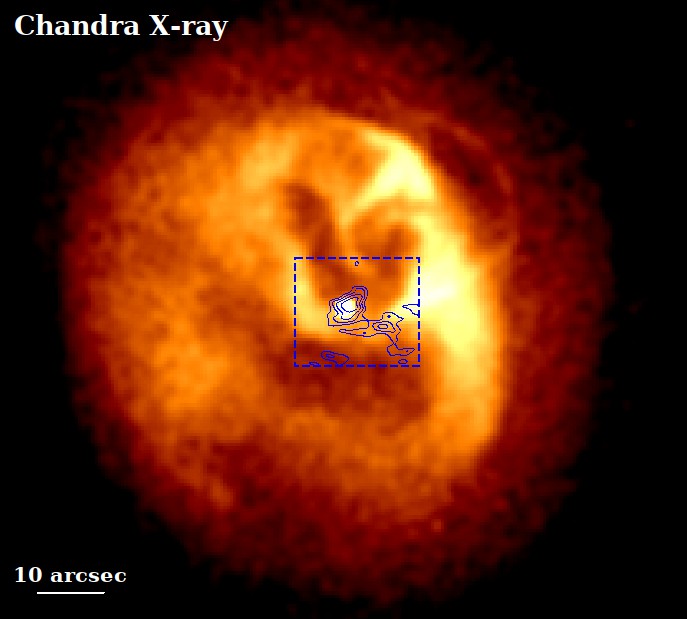}}
\includegraphics[width=0.35\columnwidth]{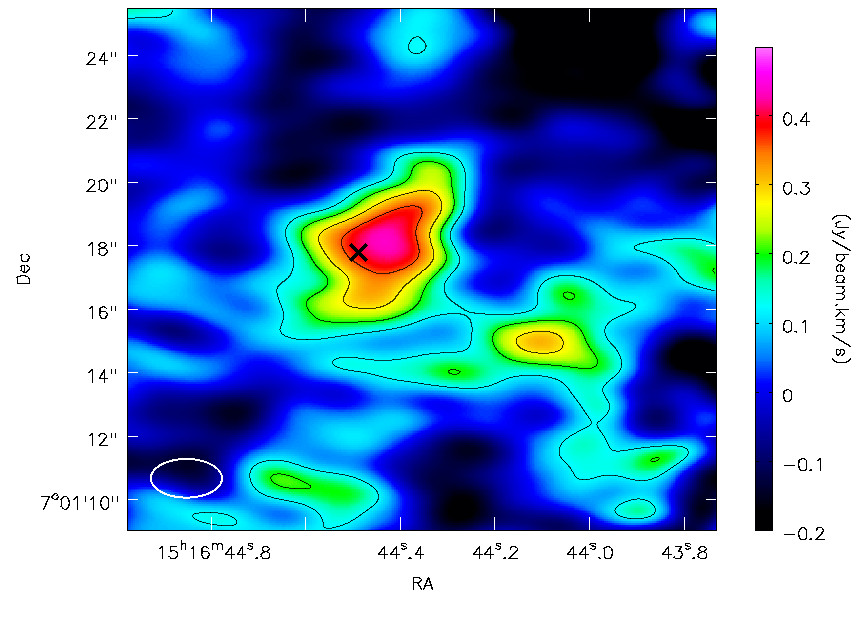}
%\raisebox{0.3cm}{\includegraphics[width=0.3\columnwidth]{A2052_hstf555wsmall.jpeg}}
\caption{Abell 2052.  Left: SDSS R-band image showing the galaxies at the cluster centre.  Centre: Chandra X-ray image showing the hot cluster atmosphere, central AGN and a series of cavities along the N-S axis.  Right: CO(2-1) integrated intensity map for the velocities $-150$ to $+100\kmps$ with contours at $-3\sigma,3\sigma,5\sigma,7\sigma ...$, where $\sigma=0.04\Jypbmkmps$.  The position of the sub-mm continuum point source is marked with a black cross and the field of view of the CO(2-1) image is shown by the red and blue boxes.}%  Note that the CO(2-1) image has been corrected for the decrease in sensitivity over the ALMA field of view to ensure accurate flux densities.  However, this overestimates the significance of features at the edges of the image (ie. apparent peaks at radii $>15\asec$).  By measuring the signal-to-noise in an uncorrected image, we estimate that the NNW and WSW features are detected at $6\sigma$, the ENE feature is detected at $4\sigma$ and the SSE feature is marginally detected at close to $3\sigma$.  Whilst the NNW feature is coincident with a particularly bright filament of H$\alpha$ emission over the N radio bubble, the other features are not coincident with any optical line emission or star formation and the symmetry of these features about the bright central peak suggests they are noise.}
\label{fig:a2052}
\end{minipage}
\end{figure*}

\begin{figure*}
\begin{minipage}{\textwidth}
  \centering
\raisebox{0.3cm}{\includegraphics[width=0.29\columnwidth]{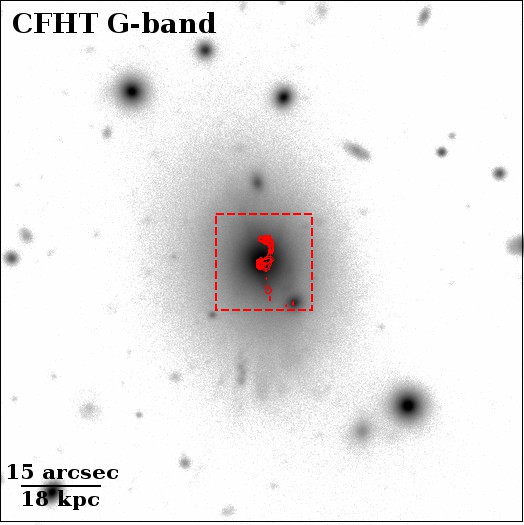}}
\raisebox{0.3cm}{\includegraphics[width=0.29\columnwidth]{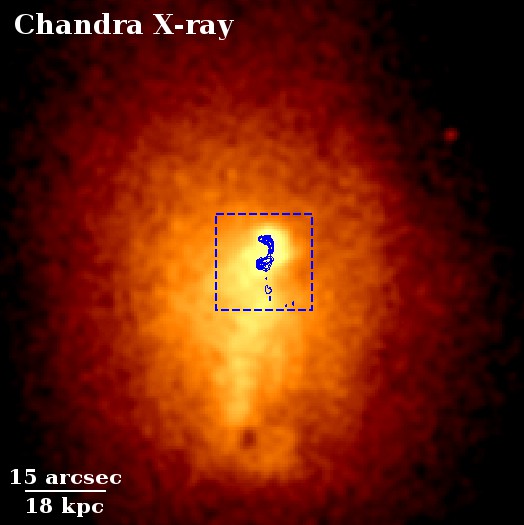}}
\includegraphics[width=0.4\columnwidth]{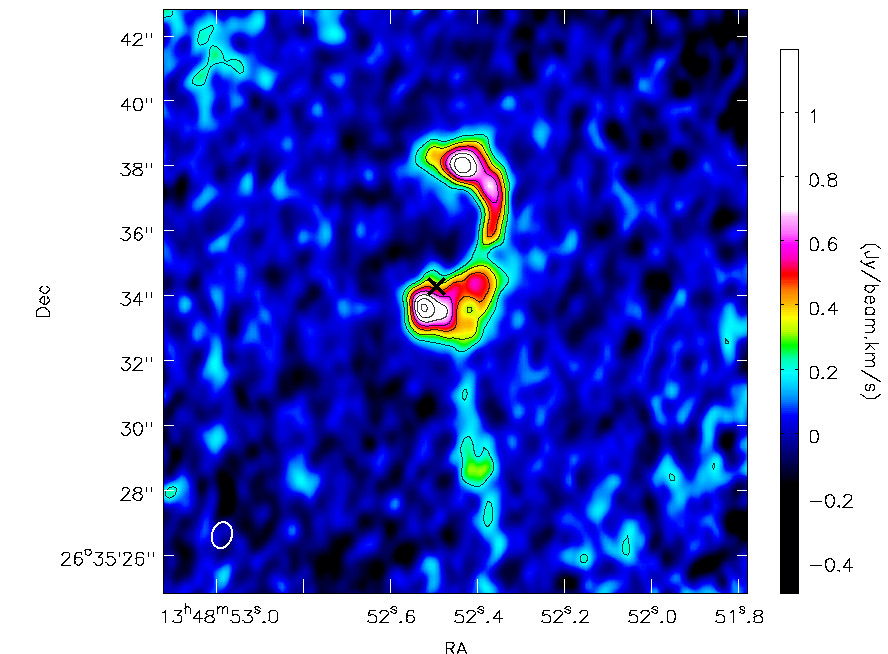}
\caption{Abell 1795.  Left: CFHT G-band archival image of the central galaxy.  Centre: Chandra X-ray image showing the hot cluster atmosphere and the $46\kpc$-long soft X-ray filament that extends S of the galaxy centre.  Right: CO(3-2) integrated intensity map for velocities $-340$ to $+130\kmps$ with contours at $-3\sigma,3\sigma,5\sigma,7\sigma ...$, where $\sigma=0.064\Jypbmkmps$.  The position of the sub-mm continuum point source is marked with a black cross and the field of view of the CO(3-2) image is shown by the red and blue boxes.  The CO(2-1) contours are also shown overlaid on the optical and X-ray images.}
% Don't show CO(1-0)? essentially consistent with CO(3-2) ...
\label{fig:a1795}
\end{minipage}
\end{figure*}

The filaments are typically a few kpc in length on the sky but can
extend up to $10-20\kpc$ (eg. Phoenix and Perseus).  The measured
filament extent will be dependent on the depth of each observation and
the maximum resolvable scale of the array configuration (typically
$5-10\asec$ for these observations).  Whilst comparisons between
single dish and ALMA flux measurements suggest that the bulk of the
molecular gas is captured by these ALMA observations
(eg. \citealt{Russell14,McNamara14,David14}), more extended, fainter
filaments may have gone undetected.  Molecular gas filaments are known
to spatially coincide with ionized and soft X-ray filaments
(\citealt{Salome04,Salome06,Lim08}), which typically extend
significantly beyond the filaments detected in the existing ALMA
observations (eg. A2052 Fig. \ref{fig:a2052}, A1795
Fig. \ref{fig:a1795}).  For example, CO emission coincident with
H$\alpha$-emitting filaments has been detected out to $50\kpc$ radius
in single dish observations of the Perseus cluster
(\citealt{Salome11}).  Star-forming filaments in the Phoenix cluster
extend even further to $100\kpc$ radius (\citealt{McDonald15}).
Therefore, whilst deeper observations with more compact ALMA
configurations will likely reveal fainter, more extended structure,
the majority of the molecular gas mass lies in filaments that are a few to
$15\kpc$ long.

% The fraction of the molecular gas reservoir that lies in the filaments
% ranges from $10-20\%$ (A1835, A262) to $>70\%$ (A1795, PKS\,0745).  In
% PKS\,0745 and A1795, the molecular gas lies exclusively in extended
% filaments drawn up around or extending towards X-ray cavities
% (Figs. \ref{fig:pks07} and \ref{fig:a1795}).  Whilst in A262 the
% majority of the molecular gas is resolved in a rotating circumnuclear
% gas disk and only a small fraction extends to larger radii along the
% jet axis (Fig. \ref{fig:A262maps}).  These numbers are lower limits as
% the limited spatial resolution in several cases will produce
% underestimates.  For example, the Early Science observations of A1835
% do not resolve the majority of the molecular gas, which lies in the
% central peak.

% filaments due to patchy shells?

\begin{figure*}
\begin{minipage}{\textwidth}
\centering
\includegraphics[width=0.32\columnwidth]{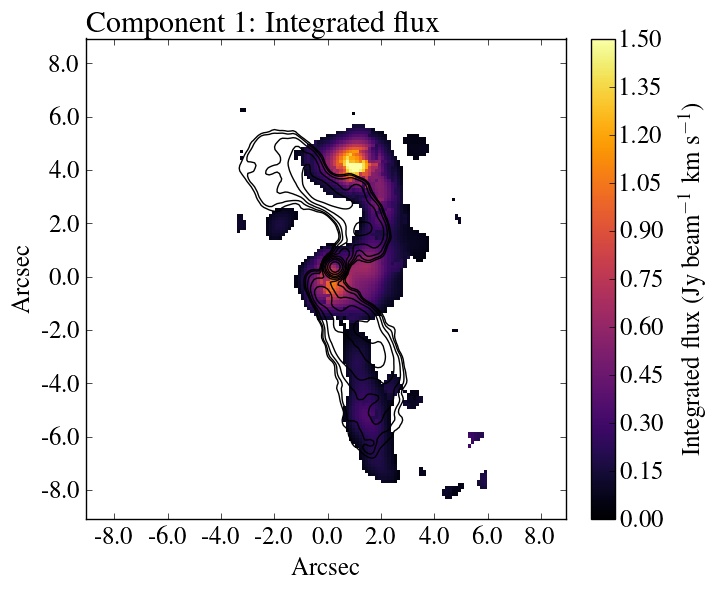}
\includegraphics[width=0.33\columnwidth]{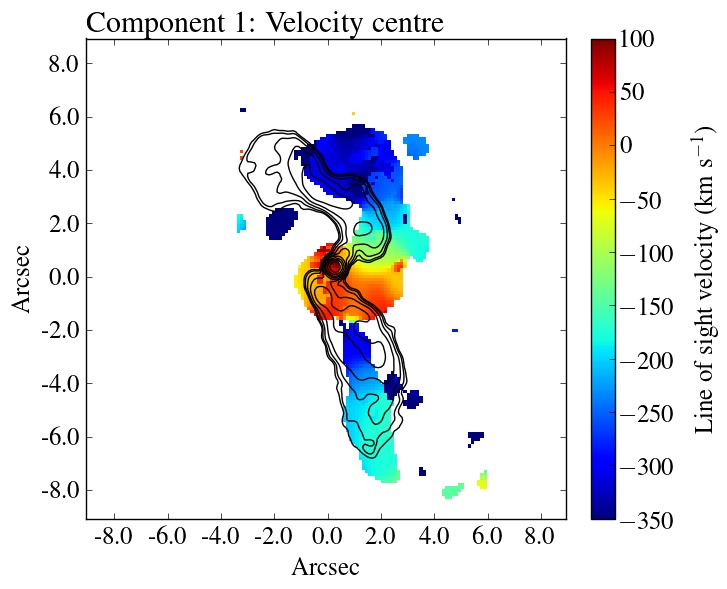}
\includegraphics[width=0.32\columnwidth]{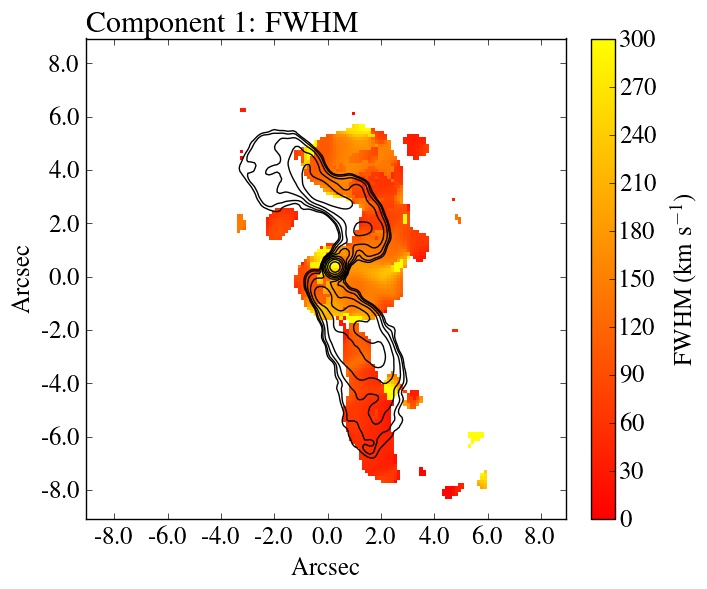}
\includegraphics[width=0.32\columnwidth]{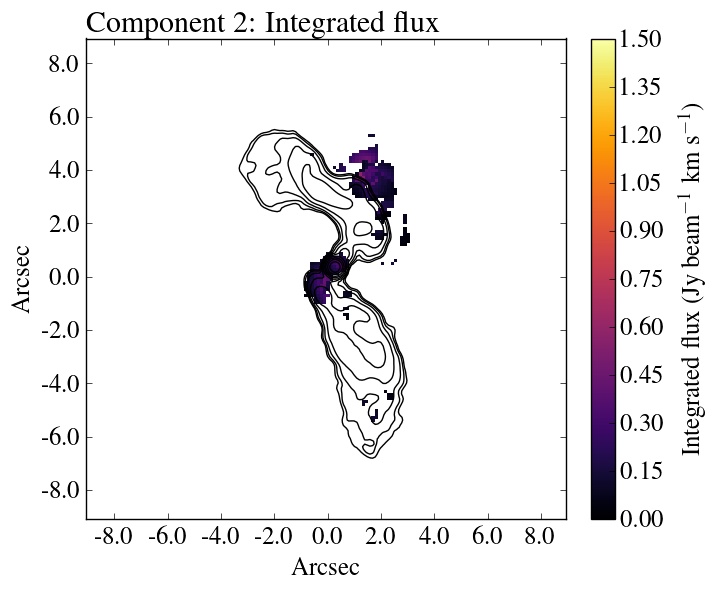}
\includegraphics[width=0.33\columnwidth]{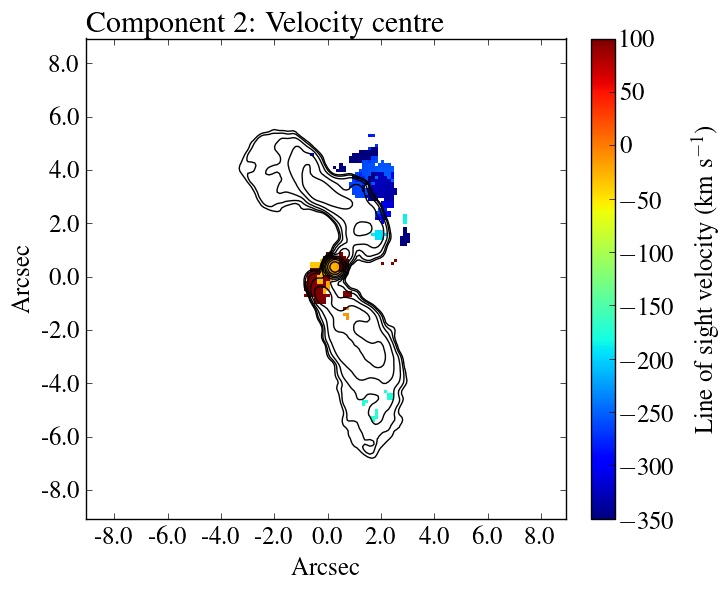}
\includegraphics[width=0.32\columnwidth]{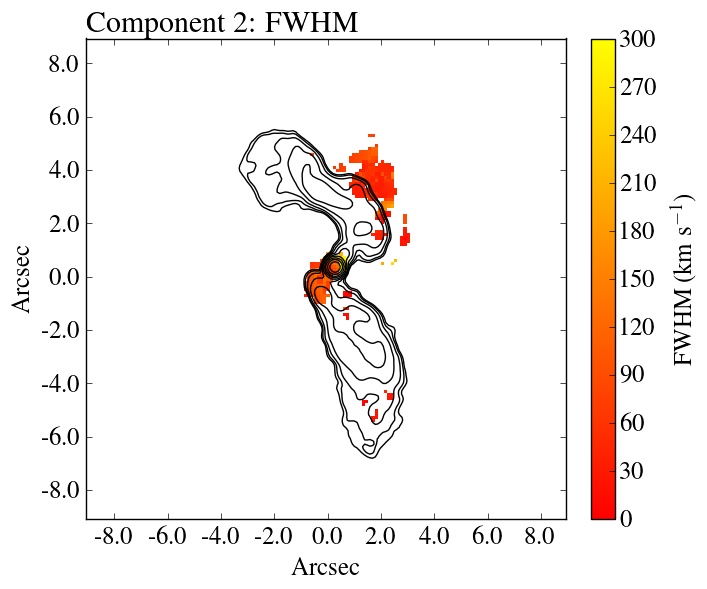}
\caption{Maps of the best-fit integrated intensity (left), velocity centre (centre) and FWHM (right) in A1795 for Gaussian components detected at $>3\sigma$.  VLA $4.9\GHz$ contours are superimposed (\citealt{vanBreugel84}).}
\label{fig:a1795maps}
\end{minipage}
\end{figure*}

\begin{figure*}
\begin{minipage}{\textwidth}
  \centering
\includegraphics[width=0.27\columnwidth]{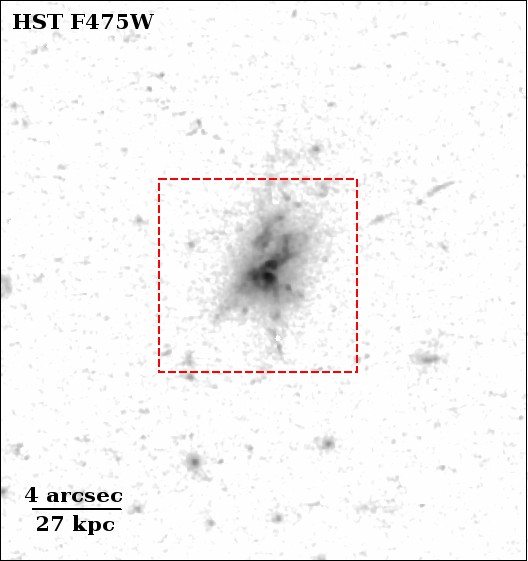}
\includegraphics[width=0.27\columnwidth]{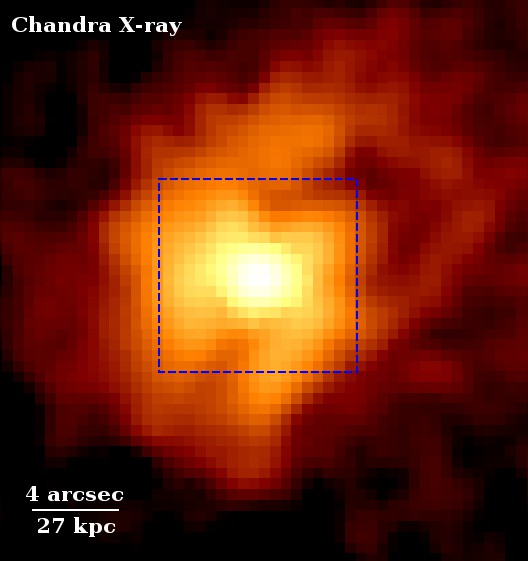}
\includegraphics[width=0.4\columnwidth]{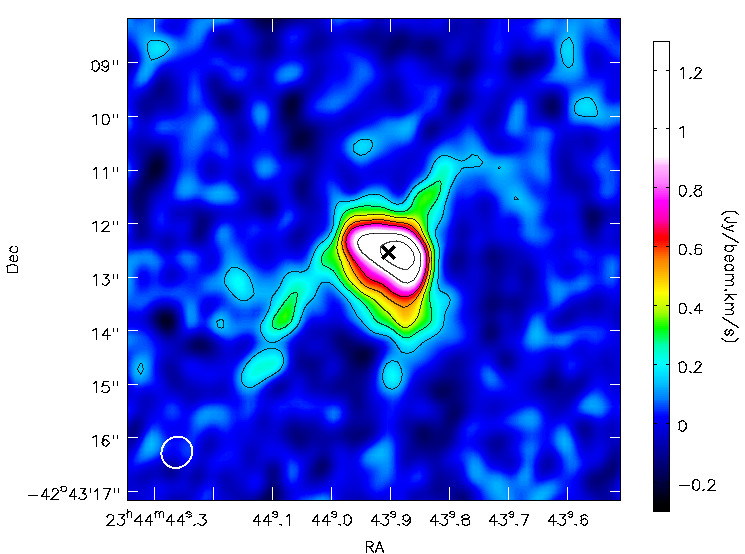}
\caption{Phoenix cluster.  Left: HST F475W archival image of the central galaxy.  Centre: Chandra X-ray image showing the hot cluster atmosphere.  Right: CO(3-2) integrated intensity map for velocities $-430$ to $+600\kmps$ with contours at $2\sigma,4\sigma,6\sigma,8\sigma,10\sigma,15\sigma,20\sigma ...$, where $\sigma=0.067\Jypbmkmps$ (from \citealt{Russell17}).}
\label{fig:phoenix}
\end{minipage}
\end{figure*}

A clumpy, thin shell of molecular gas that surrounds an X-ray cavity
will appear brightest around the rim where the line of sight through
this gas is greatest.  This would explain why the molecular gas is preferentially detected as
filaments around the outer edges of the X-ray cavities in several
observations.  However, this does not explain why molecular gas is preferentially
detected along only the outer edge of each radio lobe in A1795 and the
W side of the N X-ray cavity in the Phoenix cluster
(Figs. \ref{fig:a1795} and \ref{fig:phoenix}).  This uneven
distribution could be due to the clumpy nature of the filaments or the
collapse of some molecular filaments into stars (eg. A1795,
\citealt{Russell17A1795}).  A particularly bright molecular gas clump
in the N filament of A1795 is also spatially coincident with a notch
in the radio lobe.  The additional molecular velocity components in
this region (Fig. \ref{fig:a1795maps}), together with increases in the
ionized gas velocity, line width and ionization state
(\citealt{Crawford05}), suggest possible collisions with the expanding
radio lobe.  The collision may enhance emission from the molecular gas.

% offsets in peak from nucleus

\begin{figure*}
\begin{minipage}{\textwidth}
  \centering
\raisebox{0.5cm}{\includegraphics[width=0.29\columnwidth]{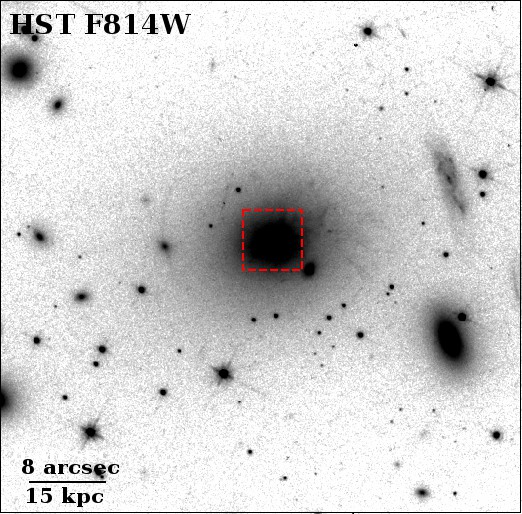}}
\raisebox{0.5cm}{\includegraphics[width=0.29\columnwidth]{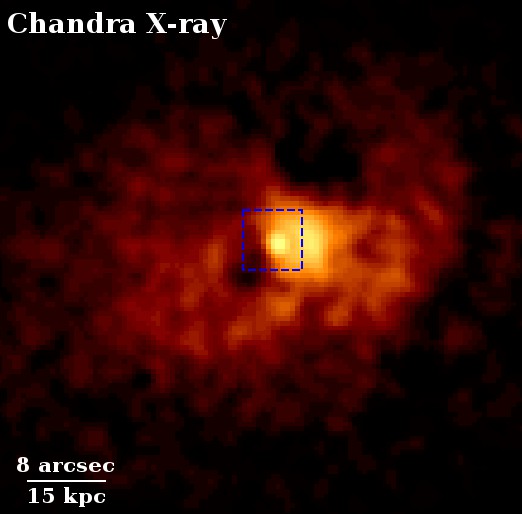}}
\includegraphics[width=0.4\columnwidth]{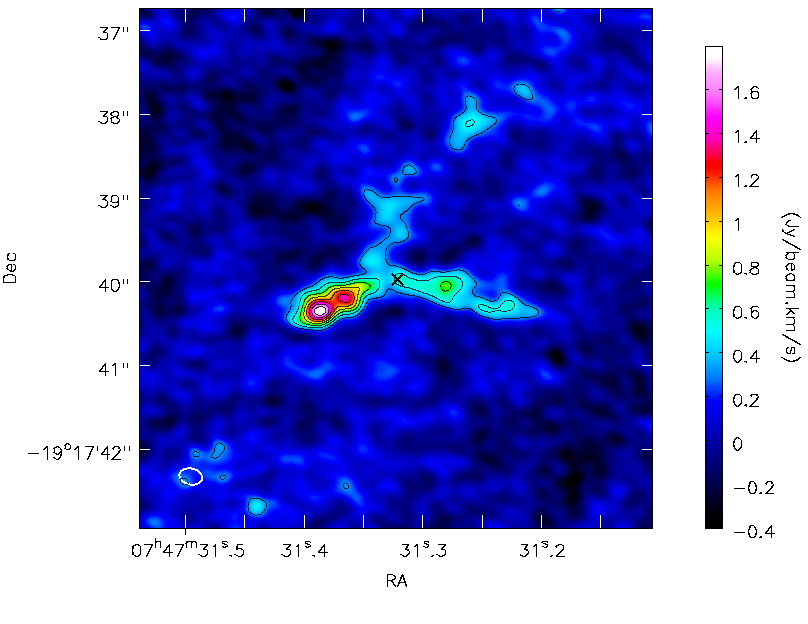}
\caption{PKS0745-191.  Left: HST F814W image showing the galaxies at the cluster centre.  Centre: Chandra X-ray image showing the hot cluster atmosphere, the bright central AGN and two cavities to the NW and SE of the nucleus.  Right: CO(3-2) integrated intensity map for velocities $-240$ to $+180\kmps$ with contours at $-3\sigma,3\sigma,5\sigma,7\sigma ...$, where $\sigma=0.1\Jypbmkmps$.  The position of the sub-mm continuum point source is marked with a black cross and the field of view of the CO(3-2) image is shown by the red and blue boxes.  The CO(1-0) image shows similar structure but at lower spatial resolution (\citealt{Russell16}).}
% Don't show CO(1-0)? essentially consistent with CO(3-2) ...
\label{fig:pks07}
\end{minipage}
\end{figure*}

In addition to the prevalence of extended molecular filaments, the
molecular emission peak in the majority of the ALMA targets is offset
from the AGN by projected distances of $\sim0.5-3\kpc$.  The
exceptions are the disk-dominated systems, such as A1664 and A262,
where the nucleus appears spatially coincident with the molecular peak
within the uncertainties.  RXJ0821 features the largest projected offset of
$\sim3\kpc$ between the central galaxy's nucleus and the molecular gas
peak.  \citet{Vantyghem18} suggest the wholesale displacement of the
$\sim10^{10}\Msun$ molecular reservoir in this system is due to
sloshing motions in the intracluster medium triggered by
the close passage of a nearby galaxy.  In the rest of the sample,
the offsets between the AGN and the molecular peak are more typically
$1-2\kpc$ and in several systems, where the filamentary structure is
resolved, this is due to bright gas clumps in the filaments entrained
around X-ray cavities (eg. PKS\,0745, A1795, Figs. \ref{fig:pks07} and
\ref{fig:a1795}).

%%% Insert punchy summary

In summary, the cold gas morphology in our sample of central galaxies
ranges between filament-dominated systems, including Perseus and
A1795, and rarer disk-dominated systems, such as Hydra A and A262.  In
filament-dominated systems, over 70\% of the molecular gas lies in
several massive filaments that extend radially out to $10\kpc$.  For
the majority of the central galaxies targeted, the molecular gas is
split more evenly between the filamentary structure and a
circumnuclear peak.  Molecular filaments are entrained around or extended
toward radio bubbles or X-ray cavities in at least 50\% of the systems observed.  Deeper
observations with improved spatial resolution will likely increase
this fraction.  Disk-dominated systems also appear rare in our sample.
Although we are limited by small number statistics, the ALMA targets
were not selected on their dynamics so the low fraction of disks is
likely to be representative.

\subsection{Velocity structure}
\label{sec:vel}

\begin{figure*}
\begin{minipage}{\textwidth}
\centering
\includegraphics[width=0.45\columnwidth]{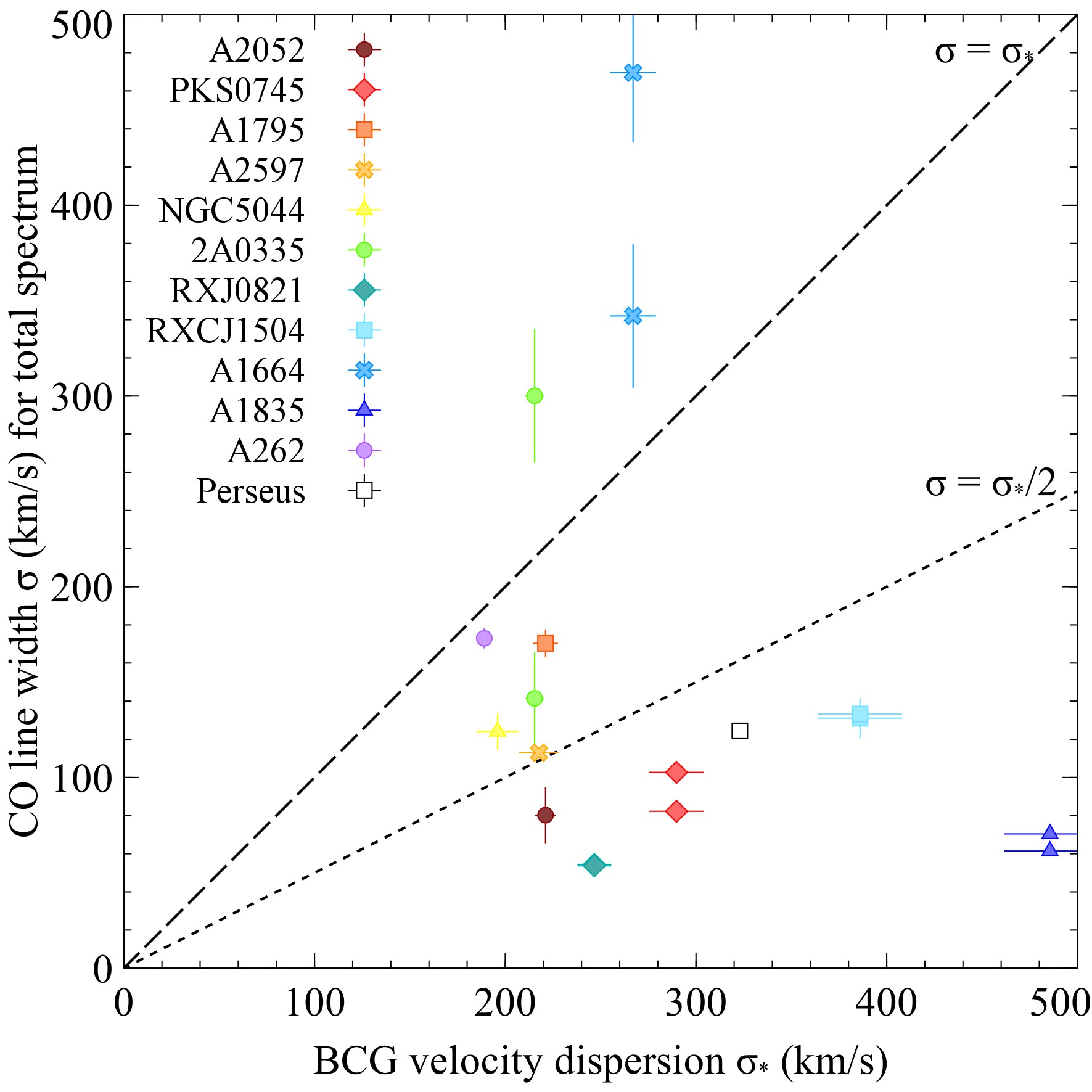}
\includegraphics[width=0.45\columnwidth]{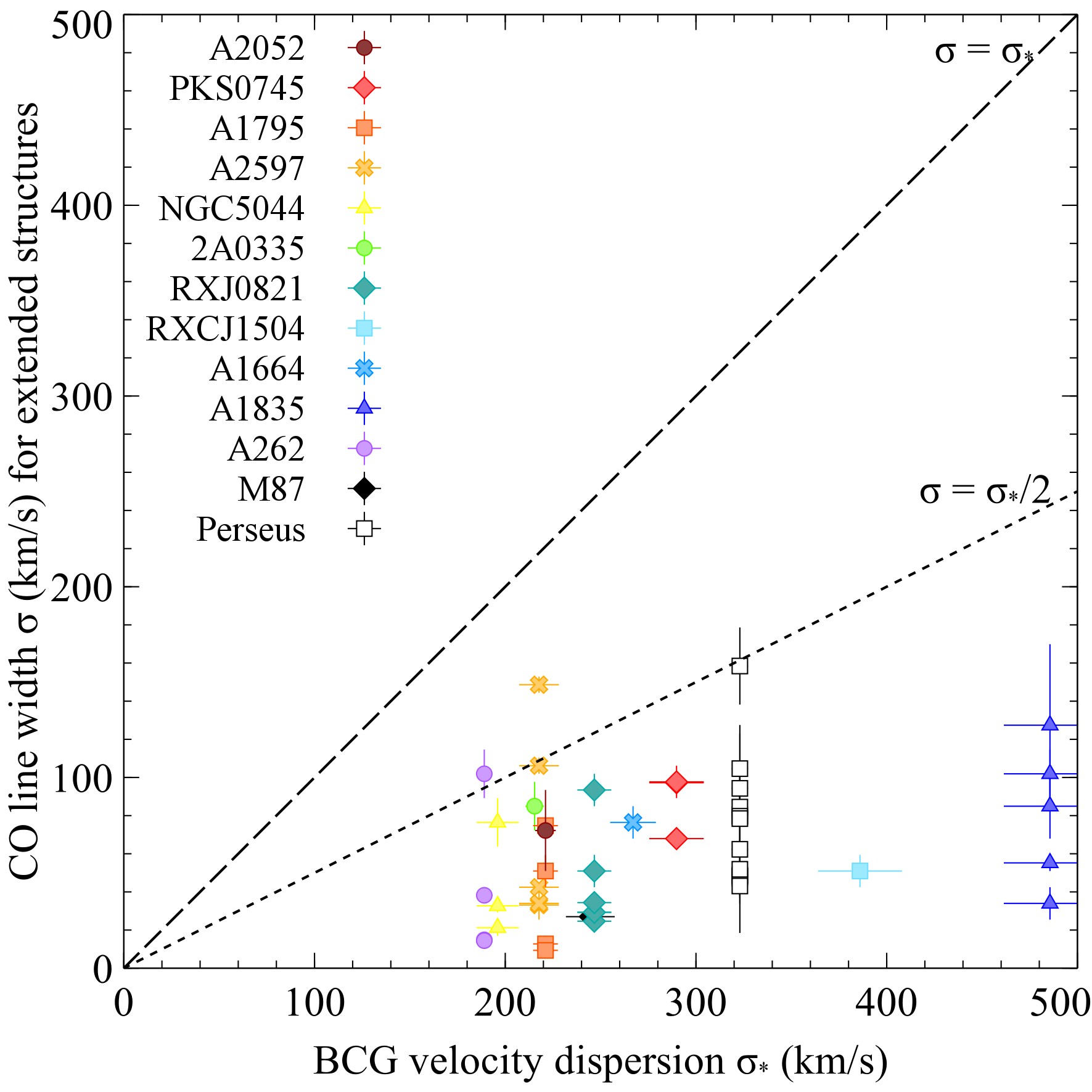}
\caption{Equivalent stellar velocity dispersion of the central galaxy from \citet{Hogan17} and \citet{Pulido18} vs the CO line width (see appendix) for a single Gaussian fit to the total spectrum in each CO line (left) and individual velocity components in each extended molecular structure (right).  Several sources were observed in both CO(1-0) and CO(3-2) and therefore have two data points.  Note that the Phoenix cluster does not have a measurement of stellar velocity dispersion in the literature.  For M87 and Perseus, we utilize ALMA and IRAM $30\m$ CO(2-1) results from \citet{Simionescu18} and \citet{Salome11}, respectively, and stellar velocity dispersions from \citet{Ho09Disp}.}
\label{fig:disp}
\end{minipage}
\end{figure*}

Fig. \ref{fig:montagevel} shows a subset of the velocity maps to
demonstrate the smooth velocity gradients along the extended filaments
drawn up around radio bubbles and the ordered rotation in the
circumnuclear disks.  In the Perseus cluster, several extended ionized
gas filaments have smooth velocity gradients along their lengths,
which match molecular gas velocities in overlapping regions from
single dish observations (\citealt{HatchPer06,Revaz08,Salome11}).  These
gradients match simulations of streamlines behind buoyantly rising
radio bubbles.  We note that the smaller strands that make up these filaments may have more complex dynamics within this larger scale smooth gradient (eg. \citealt{HatchPer06,GendronMarsolais18}).

Similarly, ALMA observations of the molecular filaments around or
beneath radio bubbles in A1795, Phoenix, PKS\,0745, 2A\,0335,
RXCJ\,1504 and RXJ\,0821 have smooth velocity gradients along their
lengths or over few kpc-long sections.  A1795 and Phoenix represent
the most spectacular examples where filaments are exclusively drawn up
around radio bubbles.  The gas velocities are remarkably ordered over
$5-15\kpc$ in length and several hundred $\kmps$.  In A2052, the
velocity of the molecular gas blobs is consistent with the
spatially-coincident ionized filaments, which also have smooth
gradients around the radio bubble rims (\citealt{Balmaverde18}).

The disk-dominated systems also feature an ordered velocity structure
consistent with rotation about the galaxy centre.  The archetype Hydra
A hosts an edge-on disk, $\sim5\kpc$ across.  IFU observations of the
ionized and warm molecular hydrogen show rotation in a plane
perpendicular to the radio jet axis (\citealt{Hamer14}).  ALMA
observations of A262 show a similarly ordered circumnuclear disk from
$-200$ to $+200\kmps$ centred on the AGN and oriented perpendicular to
the radio jet axis (Fig. \ref{fig:A262maps}).  The disk's velocity structure matches that observed in the ionized gas (\citealt{Hatch07}) and is consistent with the IRAM $30\m$ CO line profile (\citealt{Prandoni07}).  A1664 hosts two distinct velocity
structures: a high velocity gas flow at $-600\kmps$ and a possible
nascent disk spanning $-200$ to $+200\kmps$ centred on the nucleus
(\citealt{Russell14}).  Similarly, in addition to the extended
filaments, the Phoenix cluster has a second velocity component in the
circumnuclear peak which exhibits a smooth velocity shift from $-200$
to $+200\kmps$ across the nucleus over $\sim7\kpc$
(\citealt{Russell17}).  This could correspond to a rotating gas disk
that is oriented perpendicular to the radio bubble axis.  Other
systems, such as A1835 and A2597, host more complex circumnuclear
structure that is suggestive of rotation but is currently poorly
resolved (\citealt{McNamara14,Tremblay18}).

%% Summary

In summary, both the circumnuclear disks and filaments
entrained around radio bubbles exhibit ordered velocity structure.
The filaments have smooth, shallow velocity gradients spanning a few
hundred $\kmps$ and a few to tens of kpc.  Simulations have shown that
these velocity gradients are consistent with gas flows tracing
streamlines around the buoyantly rising radio bubbles.  Circumnuclear
gas disks are rarer in our sample.  The clear examples in A262 and
Hydra A show ordered rotation over $\sim3-5\kpc$ and $\sim500\kmps$ in a
plane perpendicular to the radio jet or lobe axis.

\subsubsection{Low molecular gas velocities}
As has previously been noted for individual ALMA targets
(\citealt{McNamara14,Russell16}), the molecular gas velocities in
these systems are surprisingly low and generally fall significantly
below the stellar velocity dispersion.  The molecular gas
  velocities are also much lower than the escape velocity for the
  central galaxies in this sample, typically $\sim1000\kmps$.
Therefore the molecular gas remains firmly bound to the galaxy even in
the highest velocity structures, such as the high velocity filament in
A1664 at $-600\kmps$ (Fig. \ref{fig:A1664maps}).

Fig. \ref{fig:disp} (left) compares the equivalent stellar velocity
dispersion of the central galaxy ($\sigma_*$, \citealt{Hogan17,Pulido18}) with the
CO velocity dispersion ($\sigma$) for a single Gaussian component fit to each
target.  The CO line widths are particularly narrow compared to the
stellar dispersions for the majority of our sample with
$\sigma_{*}>\sigma$.  The exceptions are A262, which is dominated by a
rotating gas disk, $\sim3\kpc$ across, and A1664 and 2A0335, which feature multiple
velocity components spread over a wide range in velocity and are
therefore particularly poorly described by a single Gaussian model.
In A1835, RXCJ1504, PKS0745, RXJ0821 and A2052, the CO line width
falls significantly below half of the equivalent stellar velocity
dispersion.  Comparisons with single dish detections suggest that
these ALMA observations detect the majority of the CO line emission
and with similar line widths, which indicates that significant broader
velocity components have not been missed.  Therefore, for at least
$75\%$ of the targets analysed, the cold molecular gas is not
dynamically relaxed in the central galaxy's gravitational potential.

Fig. \ref{fig:disp} (right) demonstrates that the extended filaments
are even more extreme.  The CO velocity dispersion measured for the
vast majority of the filaments falls significantly below half of the
equivalent stellar velocity dispersion.  The filaments are not
rotationally supported.  Unless supported by another mechanism, they
should free-fall in the cluster's deep gravitational potential.  Using
models for the cluster potential, previous studies have shown that
free-falling gas blobs in these systems will be accelerated to
velocities of at least a few hundred $\kmps$ over distances of only a
few kpc (eg. \citealt{Lim08,Russell16,Russell17,Vantyghem16}).  The
observed velocity gradients along the filaments are therefore much too
shallow compared to the predictions for free-fall.  Unless all
filaments in the sample are oriented within $\sim20\deg$ of the plane
of the sky, the velocity gradients of the most extended filaments are
generally inconsistent with free-fall.

We note that observations at higher spatial resolutions, particularly
of the more distant sources Phoenix and A1835, may reveal strands
within each filament with more complex velocity structure.  Whilst
single-dish observations of Perseus at spatial resolutions of several
kpc showed no overall pattern in the gas kinematics, SMA
interferometric observations at a spatial resolution of $\sim1\kpc$
revealed three radial filaments, of which the outer two have a
velocity structure consistent with free fall (\citealt{Lim08,Ho09}).
The majority of the targets in our sample are observed with spatial
resolutions of a kpc or better in at least one CO emission line
(eg. Table \ref{tab:cube}).  However, observations of Phoenix and
A1835 currently have spatial resolutions of a few to $5\kpc$.
Therefore, similar to Perseus, more detailed observations may reveal
more complex velocity structure.

\subsubsection{Inflow and outflow}
\label{sec:inflowoutflow}

% Not possible to distinguish inflow from outflow

Smooth, radial velocity gradients along the filaments, combined with
the clear morphological alignments with X-ray cavities and radio bubbles, suggest that
the filaments are gas flows either inflowing or outflowing in the
bubbles' wakes.  Unless the gas clouds in the filaments are
also detected in absorption, it is impossible to determine whether
they are located on the near or far-side of the central galaxy with
respect to the nucleus.  An absorption signature against the continuum
emission from the AGN would place the gas cloud on the near-side of
the galaxy, in front of the AGN.  A blueshifted line would then
indicate an outflow and a redshifted line would indicate an inflow.
Absorption lines have been detected against the sub-mm nuclear
continuum emission in NGC\,5044, A2597 and Hydra A
(\citealt{David14,Tremblay16,Rose19}).  The absorbing clouds have
similarly low velocities as the emitting clouds and cover the
same narrow dynamical range.  The apparent motion relative to the AGN
may indicate that these clouds are inflowing
(\citealt{David14,Tremblay16}) or on stable, low ellipticity orbits
(\citealt{Rose19}).  In the absence of absorption lines for the vast
majority of the observed structures, general conclusions can be drawn
from a sample of these filaments.

% However similarity in velocity at large radius in Phoenix, and A1795 gas appears to park itself at the BCG velocity

Filament-dominated systems are difficult to understand in a pure
outflow scenario where molecular clouds are directly lifted from the
galaxy centre by the radio bubbles.  The radio bubbles would have to
efficiently couple to the molecular gas to draw such a large fraction
of the gas, in some cases exceeding $70\%$ (section
\ref{sec:morphology}), into extended filaments.  This must be
maintained over large distances out to $30\kpc$ to explain the
observed filament lengths.  The coupling must also be gentle.  The gas
velocities in the filaments do not typically exceed a few hundred
$\kmps$.  This is consistent with the lack of strong shocks around the
radio bubbles in the hot X-ray atmosphere.  The bubbles expand
approximately in pressure equilibrium with the surrounding
intracluster medium and the gas rims around them are relatively cool
rather than shock-heated (eg. \citealt{McNamara00,FabianPer00}).  The
morphology and kinematics of these gas flows are therefore starkly
different from the fast ($>500\kmps$) molecular outflows in nearby
Seyferts, which are directly accelerated by interactions between the
relativistic jet and the circumnuclear gas disk
(eg. \citealt{Morganti13,Garcia-Burillo14}; Tamhane et al. in prep.).

The remarkably large lifted fractions of molecular gas in the
filament-dominated systems could be explained if the cold filaments
originate in rapid cooling from a hot gas flow (\citealt{McNamara14}).
X-ray observations show that the radio bubbles displace and lift a
substantial mass of the low entropy X-ray gas surrounding the central
galaxy.  These hot gas flows can also be clearly traced as metal-rich
plumes of gas along the radio bubble axis in nearby systems
(\citealt{Simionescu08,Kirkpatrick09}) and are a feature of
hydrodynamic simulations of AGN feedback in clusters
(eg. \citealt{Pope10,Gaspari11}).  Cool gas nebulae and star formation
are preferentially detected in central cluster galaxies when the
entropy index in the hot atmosphere falls below $\sim30\keVcmsq$
(\citealt{Cavagnolo08,Rafferty08}).  This sharp threshold implies that
the cold gas originates from the development of thermal instabilities
in the hot atmosphere
(\citealt{Nulsen86,PizzolatoSoker05,Gaspari13,Voit15}), which are
stimulated when low entropy gas is lifted in the wakes of radio
bubbles (\citealt{Salome06,Salome11,McNamara16}).  The cool gas would
then trace streamlines around and behind the radio bubble, similar to
the observed radial filamentary morphology, and be spatially
coincident with filaments of soft X-ray emission and intermediate
temperature gas, as observed
(\citealt{FabianPer03,HatchPer06,McDonald12A1795,Lim12}).  Hitomi
X-ray microcalorimeter observations of the Perseus cluster also showed
that the intracluster medium in the wake of the NW radio bubble has a
similar velocity gradient and low dispersion to the spatially
coincident cool gas filaments (\citealt{Hitomi16short}).  We consider
the formation, energetics and fate of these molecular flows in detail
in section \ref{sec:discussion}.

The molecular gas may retain the velocity structure of the hot gas
flow or decouple from the hot flow and fall back towards the galaxy
centre.  Molecular gas that is still coupled or recently
decoupled from a rising radio bubble may not have yet reached a high
infall speed.  In A1795, the gas velocity along the N filament
transitions smoothly from the average velocity of the surrounding
galaxies at the furthest extent to the central galaxy's systemic
velocity at the nucleus.  Similarly in Phoenix, the remarkably similar
gas velocity at the furthest extent of the filaments (regions that are $30\kpc$ apart)
suggest that the cold gas is coupled or recently decoupled from the
hot atmosphere, which is moving relative to the central galaxy.
Unless all molecular gas blobs are decoupling simultaneously along the
filament, which seems unlikely in our range of targets, we would
expect to detect higher infall velocities at small radii.  These
higher velocity gas blobs are more likely to be superimposed on other
structures at the galaxy centre and therefore potentially more
difficult to disentangle.  However, if the molecular gas was
free-falling, we would still expect to detect the corresponding higher
velocities or broader FWHM at small radii.  We would also expect to detect many more circumnuclear disks, which would grow rapidly if fed by free-falling cold gas blobs.  Only a very limited fraction of the massive molecular filaments can be consumed by the observed low levels of star formation and black hole activity.  Instead, the filaments
must be slowed and supported by an additional mechanism.

For typical ICM densities and average molecular filament densities of
$1-10\pcmcu$, ram pressure from the intracluster medium does not
significantly affect the infalling velocity of the filaments unless
the gas blobs are mists of smaller clouds (see section
\ref{sec:discbubbles}, eg. \citealt{Nulsen86,Li18}).  These structures
would have a lower mean density and would be slowed by drag in the hot
atmosphere.  Observations at higher spatial resolution could resolve
the filaments in the nearest targets to determine if they are
thread-like or fluffier clouds (eg. \citealt{Fabian08}).

Based on the survival of the extended, dense gas filaments in the
nearby Perseus cluster for at least a dynamical timescale (of order
$10^{7}\yr$), \citet{Fabian08} and \citet{Ho09} invoke the stabilizing
mechanisms of magnetic stresses and turbulence to insulate and support
the cold clouds in the hot, high pressure cluster atmosphere and
prevent their collapse.  For filament densities of $10\pcmcu$ and
temperatures $30\K$, the thermal pressure in the molecular gas is
$\sim10^{-4}$ times the thermal pressure of the surrounding hot
atmosphere.  The molecular gas might be supported by another phase or
partially by turbulence too, but another mechanism dominates the
pressure and prevents collapse.  Conditions are ripe for magnetic
support, especially if the molecular clouds consist of many thin
threads or mists of smaller clouds.

Our additional requirement that a supportive mechanism also slows
infall of the filaments along their lengths requires a more complex
magnetic field topology, such as helical fields.  The demands on
magnetic support are substantial for these massive molecular filaments
and can require a magnetic pressure roughly an order of magnitude
greater than the thermal pressure
(eg. \citealt{Russell16,Russell17A1795}).  It is not clear how such a
magnetic field topology would be generated, although we note that
simulations of buoyant radio bubbles also invoke helical field
topologies to preserve them against hydrodynamical instabilities
(eg. \citealt{Ruszkowski07, Bambic18}).

In summary, smooth, radial velocity gradients along the filaments and
clear morphological alignments with X-ray cavities and radio bubbles
suggest that the molecular filaments trace gas flows entrained by the
buoyantly rising bubbles.  The cold filaments likely originate from
the development of thermal instabilities in low entropy X-ray gas,
which is triggered when the gas is lifted in the wakes of the radio
bubbles.  The molecular gas clouds may decouple from the hot flow and
fall back towards the galaxy centre.  The smooth velocity gradients
along their lengths are significantly shallower than expected for
gravitational free-fall.  The gas clouds must be slowed, potentially
by magnetic stresses.

\subsubsection{Multiple velocity components}

\begin{figure*}
\begin{minipage}{\textwidth}
\centering
\includegraphics[width=0.32\columnwidth]{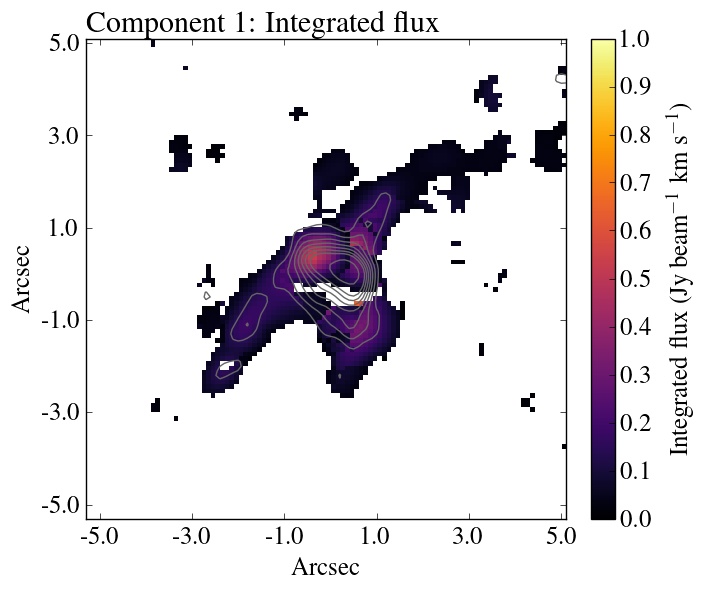}
\includegraphics[width=0.33\columnwidth]{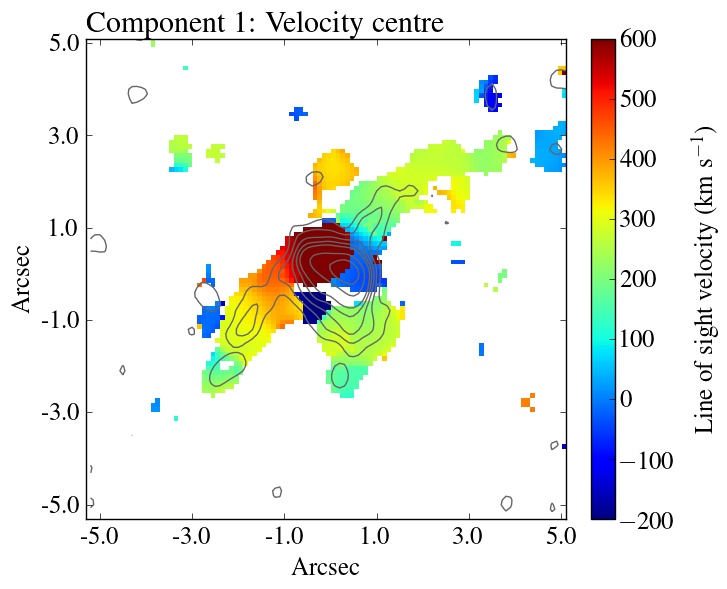}
\includegraphics[width=0.32\columnwidth]{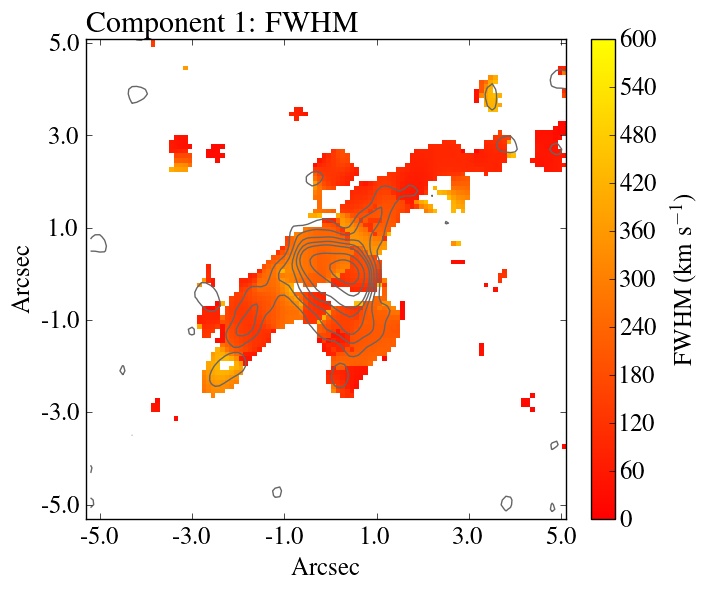}
\includegraphics[width=0.32\columnwidth]{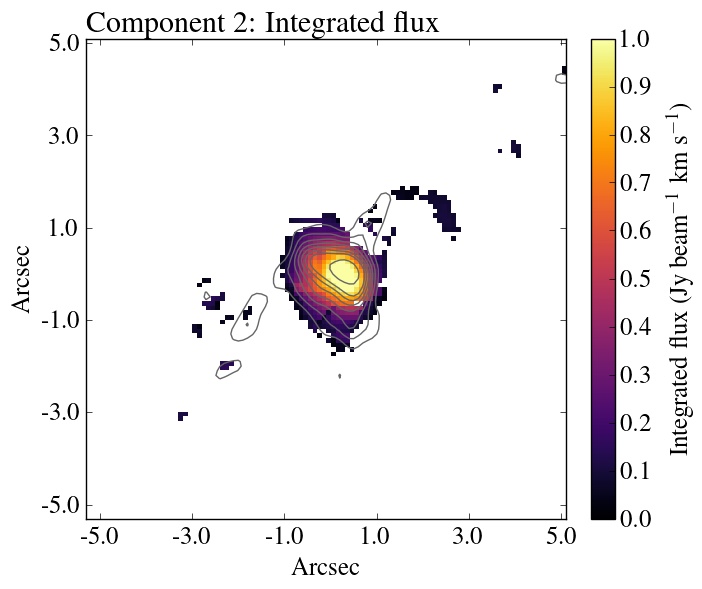}
\includegraphics[width=0.33\columnwidth]{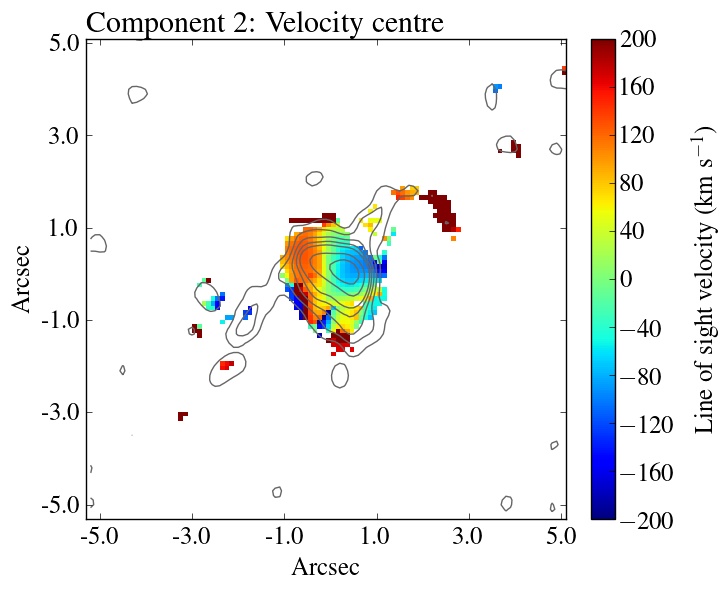}
\includegraphics[width=0.32\columnwidth]{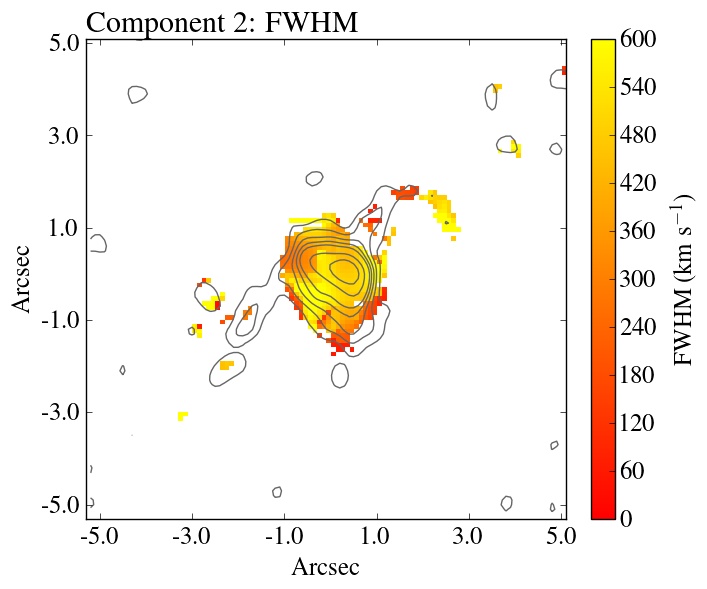}
\caption{Phoenix cluster.  Maps of the best-fit integrated intensity (left), velocity centre (centre) and FWHM (right) for Gaussian components detected at $>3\sigma$.  Integrated intensity contours from Fig. \ref{fig:phoenix} (right) are overlaid.}
\label{fig:phoenixmaps}
\end{minipage}
\end{figure*}

The velocity maps generated for this sample (section \ref{sec:vel})
also reveal additional velocity components in particular regions for
the majority of the targets analysed.  These additional components are
due to the superposition of different molecular structures along the
line of sight.  In the Phoenix cluster (Fig. \ref{fig:phoenixmaps}),
for example, the putative disk of gas with velocities from $-200$ to
$+200\kmps$ at the galaxy centre is spatially coincident with the
inner sections of the extended filaments at $+600\kmps$ and
$-200\kmps$.  These different structures that overlap in projection
can therefore be separated in velocity space.  Similarly in A1664, the
nascent disk is kinematically distinct from the high velocity
filament at $-600\kmps$ (Fig. \ref{fig:A1664maps}).

The additional velocity components may also reveal direct interactions
with the radio lobes and correspond to a superposition of infalling or outflowing filaments tracing streamlines in the wakes of bouyant bubbles (section \ref{sec:inflowoutflow}).  In A1795, A1835 and A2597,
we detect additional blue- and redshifted velocity components either
side of the nucleus that are aligned with the radio lobes and bubbles
(\citealt{Russell17A1795,McNamara14,Tremblay18}).  For A1795, these
additional components are clearly located along the outer edges of
sharp bends in the radio lobes and are coincident with increases in
the ionized gas velocity, line widths and higher ionization (\citealt{Crawford05}).  Whilst
this indicates that some molecular gas is lifted directly by the radio
jets and lobes, these
additional velocity components comprise only a small fraction of the
molecular flow.  So it seems unlikely that a large fraction of the
molecular gas is lifted in this way.  Instead, gas lifted in the hot
phase likely cools to form the bulk of the molecular clouds in situ
(see section \ref{sec:inflowoutflow}).

\begin{figure*}
\begin{minipage}{\textwidth}
\centering
\includegraphics[width=0.32\columnwidth]{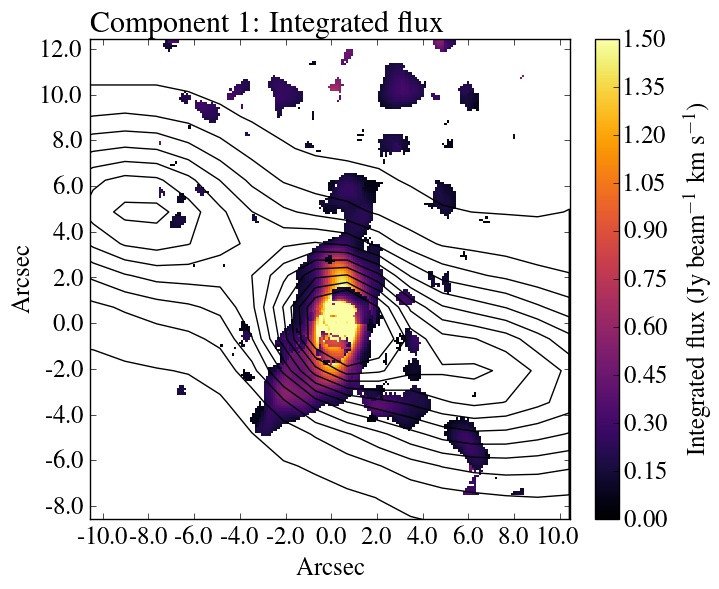}
\includegraphics[width=0.33\columnwidth]{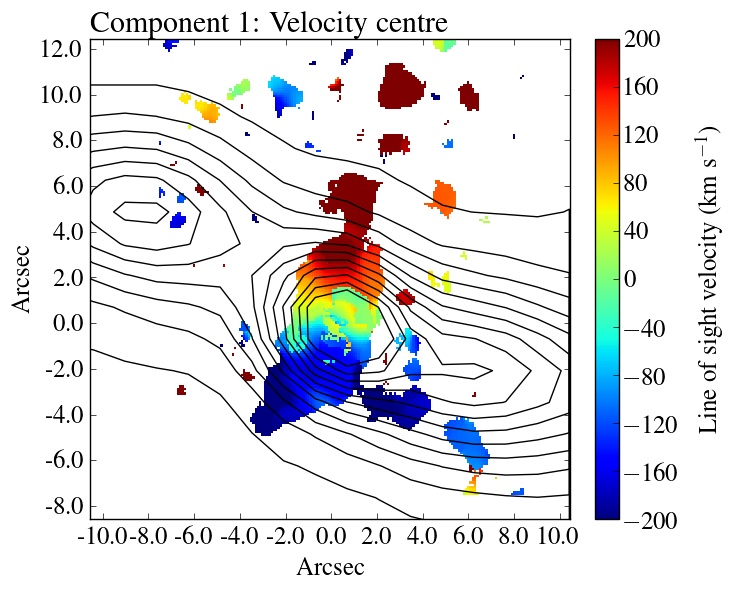}
\includegraphics[width=0.32\columnwidth]{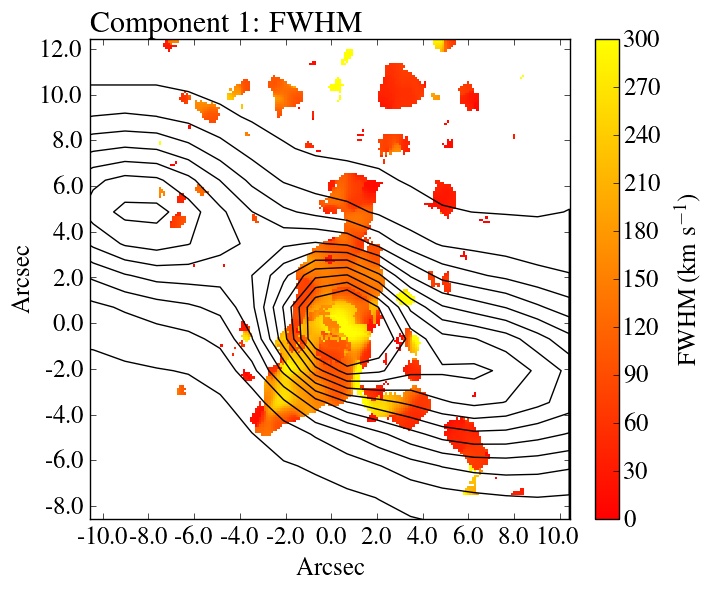}
\includegraphics[width=0.32\columnwidth]{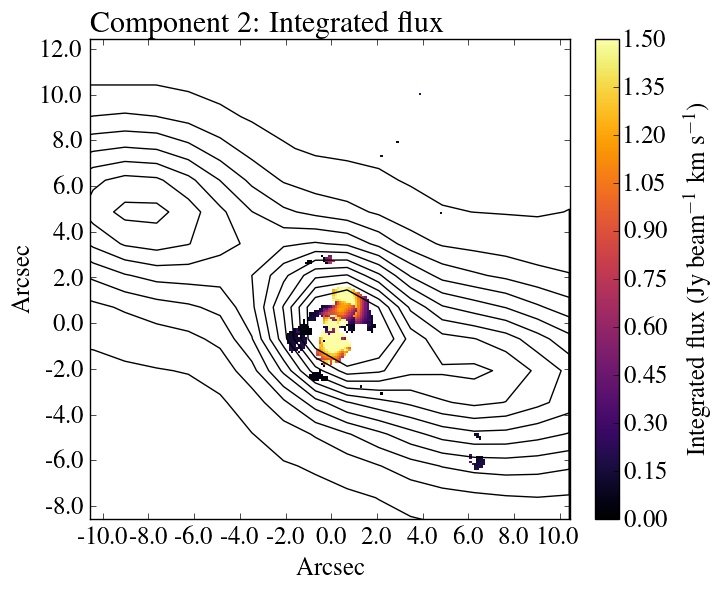}
\includegraphics[width=0.33\columnwidth]{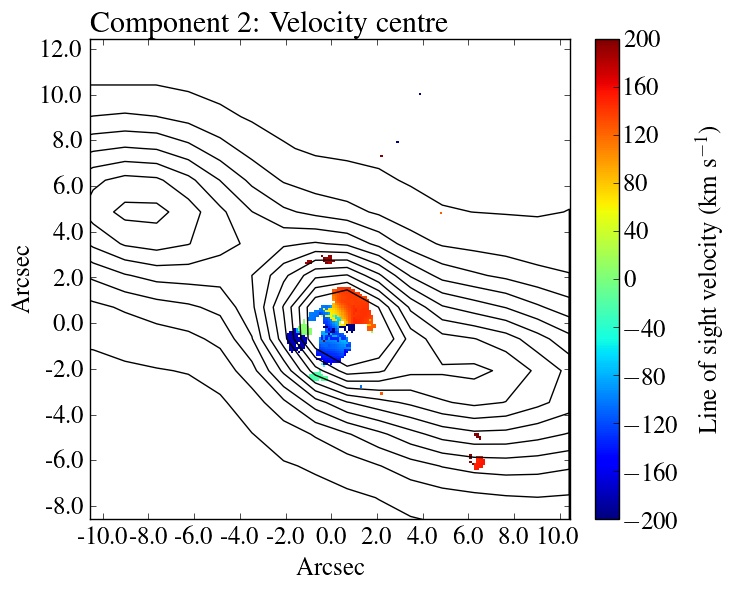}
\includegraphics[width=0.32\columnwidth]{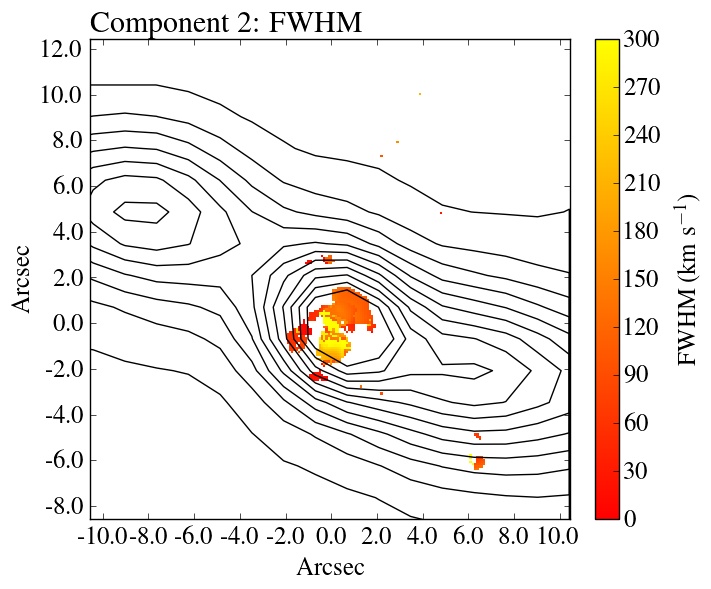}
\caption{A262.  Maps of the best-fit integrated intensity (left), velocity centre (centre) and FWHM (right) for Gaussian components detected at $>3\sigma$.  VLA $1.4\GHz$ contours are superimposed (\citealt{Clarke09}).}
\label{fig:A262maps}
\end{minipage}
\end{figure*}

Although the additional velocity component at the centre of A262
may similarly indicate a bipolar gas flow, this structure is aligned in
projection with the rotating gas disk and is oriented perpendicular to
the larger scale radio lobe axis and extended filament
(Fig. \ref{fig:A262maps}).  The radio lobes may bend through large
angles on small scales, similarly to A1795, or the additional velocity
component may be related to non-circular motions within the disk.
Similarly, for the remaining targets, including NGC\,5044, RXJ0821
and RXCJ1504, the velocity structure is
much more complex and the superimposed molecular structures overlap in
both physical and velocity space.

% eg. extra components in A2597, A1795, A262 etc.?  NGC5044, RXJ0821, RXCJ1504, Phoenix, A1664, A1835

\subsection{Line ratios}
\label{sec:lineratio}

\begin{table}
\caption{Line ratios for a subset of targets with CO(1-0) and CO(3-2) observations.}
\begin{center}
\begin{tabular}{l c c}
\hline
Target & Region & CO(3-2)/CO(1-0) \\
 & & \\
\hline
  PKS0745 & Centre & $0.80\pm0.04$ \\
       & N filament & $0.77\pm0.13$ \\
       & SE filament & $0.86\pm0.10$ \\
       & SW filament & $0.53\pm0.07$ \\
  2A0335 & Centre & $0.77\pm0.13$ \\
       & NW peak & $0.81\pm0.13$ \\
       & SE peak & $0.74\pm0.09$ \\
  RXJ0821 & E peak & $0.95\pm0.10$ \\
       & W peak & $0.98\pm0.17$ \\
  RXCJ1504 & Peak & $1.11\pm0.42$ \\
       & Inner filament & $1.18\pm0.19$ \\
       & Outer filament & $0.85\pm0.13$ \\
  \hline
\end{tabular}
\label{tab:lineratio}
\end{center}
\end{table}

% Justifies use of 0.8 as line ratio for CO(3-2)/CO(1-0) and CO(2-1)/CO(1-0).  Consistent with global values and single dish studies.  No clear case for spatial variation.

Four central galaxies in our sample were observed in detail at both
CO(1-0) and CO(3-2) and were used to measure the corresponding line
ratio.  For an optically thick medium, as expected here, the
line brightness ratio should be $\lesssim1$.  Significantly
higher CO line ratios in the extended filaments compared to the
circumnuclear gas peaks could indicate that the gas in the filaments
is more highly excited, optically thin and therefore more luminous
than the material in the disk (eg. IC\,5063, \citealt{Dasyra16}).  The
fraction of the molecular gas in the filaments would therefore be
significantly overestimated.  We note that the CO(3-2) observations
for the Cycle 0 targets, A1664 and A1835, resolve out the extended
structure traced at CO(1-0) and the global line ratio is therefore not
representative.

For the four remaining targets, PKS\,0745, 2A\,0335, RXJ\,0821 and
RXCJ\,1504, the CO(3-2) cube was convolved with an appropriate 2D
Gaussian so that the resulting synthesized beam matched that of the
CO(1-0) observation.  The integrated intensities (in $\Kkmps$) for key
molecular structures were then determined by extracting spectra from
the same spatial regions in the CO(1-0) and CO(3-2) datasets and
fitting a single Gaussian model.  The measured line ratios are
detailed in Table \ref{tab:lineratio}.

The CO(3-2)/CO(1-0) line ratio is consistent with $0.8$ for the vast
majority of the regions and targets analysed.  This is expected for a predominantly optically thick medium and in agreement with measurements of the
global line ratios in earlier single dish observations
(\citealt{Edge01,Salome03}).  There is also no clear spatial variation
in the line ratio for the extended filaments compared to the central
molecular gas peaks.

\subsection{Molecular gas mass}
\label{sec:mass}

The molecular gas mass can be estimated from the integrated CO
intensity by assuming a CO-to-H$_2$ ($X_{\mathrm{CO}}$) conversion
factor and typical brightness line ratios for BCGs of
CO(2-1)/CO(1-0)$=0.8$ and CO(3-2)/CO(1-0)$=0.8$
(eg. \citealt{Salome03}; see section \ref{sec:lineratio}).  From the integrated CO(1-0) intensity $S_{\mathrm{CO}}\Delta\nu$, the molecular gas mass is given by:

\begin{equation}
M_{\mathrm{mol}}=1.05\times10^4\left(\frac{X_{\mathrm{CO}}}{X_{\mathrm{CO,MW}}}\right)\left(\frac{1}{1+z}\right)\left(\frac{S_{\mathrm{CO}}\Delta\nu}{\Jykmps}\right)\left(\frac{D_{\mathrm{L}}}{\Mpc}\right)^2\Msun,
\end{equation}

\noindent where $z$ is the redshift of the central galaxy,
$D_{\mathrm{L}}$ is the corresponding luminosity distance and
$X_{\mathrm{CO,MW}}=2\times10^{20}\COtoH$
(eg. \citealt{Solomon87,Solomon05}).  It is not clear, however, that
the $X_{\mathrm{CO}}$ factor measured in the Milky Way and nearby
spiral galaxies is applicable to central cluster galaxies (reviewed by \citealt{Lim17}), for which
equivalent measurements are not available.  Measurements of the
$X_{\mathrm{CO}}$ factor in nearby galaxies exhibit significant
scatter and variations with environmental factors (for a review see \citealt{Bolatto13}).  

Previous studies of central cluster galaxies justified the use of
$X_{\mathrm{CO,MW}}$ by noting the approximately solar metallicity in
the surrounding ICM, line ratios indicating optically thick gas and
line widths for individual molecular clouds that are comparable to
those in the Milky Way.  The estimated factor of a
few uncertainty introduced by this approach has now been verified by \citet{Vantyghem17} who
detected both $^{12}$CO(3-2) and $^{13}$CO(3-2) in RXJ\,0821.  The
$^{13}$CO emission is generally optically thin and therefore traces
the full volume of its emitting region, which allows an estimate of
the total H$_2$ column density, molecular gas mass and
$X_{\mathrm{CO}}$ factor. \citet{Vantyghem17} showed that adopting a Galactic conversion factor could overestimate the molecular gas mass by a factor of two in RXJ\,0821.  This is easily within the object-to-object scatter from extragalactic sources.  Numerical simulations of molecular clouds with solar metallicity by \citet{Szucs16} have shown that the $^{13}$CO method of recovering the molecular gas mass systematically underpredicts the true mass by a factor of $2-3$.  This systematic would bring the estimated conversion factor in RXJ\,0821 back in line with the Galactic value.  We therefore used the Galactic CO-to-H${_2}$ conversion factor to calculate the molecular gas mass for the majority of the central galaxies in our sample with an associated factor of a few uncertainty.

The Phoenix cluster is an extreme example of a ULIRG with a star
formation rate of $500-800\Msunpyr$ (\citealt{McDonald12}).  In the
intensely star-forming environment of ULIRGs, the molecular gas exists
at higher densities and temperatures and forms an extended warm phase,
which results in far more luminous CO emission for the same gas mass
and a lower $X_{\mathrm{CO}}$ by roughly a factor of 5
(eg. \citealt{Downes93,Solomon97,Downes98}).  For the Phoenix cluster,
we therefore assume $X_{\mathrm{CO}}=0.4\times10^{20}\COtoH$ as
appropriate for a ULIRG.  As discussed in section \ref{sec:lineratio},
the observed lack of spatial variation in the CO line ratio in the
subset of galaxies analysed suggests that the physical properties of
the molecular gas are similar across the nebula.  It therefore appears
unlikely that the $X_{\mathrm{CO}}$ factor varies dramatically in the
filaments compared to the central peak.

%\subsection{Additional line detections?}
%
%Possible CN(1-0) line detection in PKS0745 dataset but not seen in
%A1835 dataset.  $^{13}$CO detection for RXJ0821.

%% Test for additional spectral lines:
% PKS0745: CN(1-0)
% A1835: covers CN(1-0) but nothing detected (odd)
% RXJ0821: nothing detected, missed CN(1-0)
% RXCJ1504: misses CN(1-0), nothing detected
% Phoenix: nothing else detected, lines missed.
% A1664: nothing else detected, should have seen 13CO(3-2) at 293.07GHz but probably too faint (typically a factor of 10 fainter so yes).

\subsection{Continuum}

For all central galaxies in our sample (except RXJ\,0821, see \citealt{Vantyghem19}), the continuum emission is unresolved and consistent with a nuclear point source.  The measured sub-mm continuum fluxes are given in Table \ref{tab:cont}.  The sub-mm continuum is typically coincident with an unresolved radio source (eg. \citealt{Hogan15}) and, for systems with deep \textit{Chandra} observations, also detected as a faint hard X-ray point source (eg. \citealt{Russell13}).  The sub-mm continuum flux is also typically consistent, within the observed variability, with synchrotron emission from the flat spectrum radio core (\citealt{Hogan15b}).  The nuclear point source therefore likely corresponds to a radiatively inefficient AGN.  The vast majority of the nuclei in these central galaxies are therefore currently accreting and active.

CO absorption lines have previously been detected in three sources
considered here, NGC\,5044, A2597 and Hydra A
(\citealt{David14,Tremblay16,Rose19}).  We confirmed these detections
but did not detect any further narrow CO absorption features against
the generally weak nuclear continuum emission in the rest of our
sample.

\begin{table}
\caption{Continuum emission for each target.}
\begin{center}
\begin{tabular}{l c c c c}
\hline
Target & $\nu_{\rm obs}$ & RA & Dec & Peak \\
 & (GHz) & & & (mJy) \\
\hline
  A2052 & 229.48 & 15:16:44.489 & +07:01:17.83 & $32.47\pm0.05$ \\
  PKS0745 & 103.53 & 07:47:31.321 & -19:17:39.97 & $8.71\pm0.02$ \\
       & 314.58 &  &  & $4.23\pm0.07$ \\ % position is consistent with CO(1-0)
  A1795 & 225.76 & 13:48:52.495 & +26:35:34.32 & $3.2\pm0.2$ \\
  A2597 & 221.33 & 23:25:19.733 & -12:07:27.18 & $14.63\pm0.02$ \\ % with self-calibration
  NGC5044 & 235.20 & 13:15:23.961 & -16:23:07.49 & $51.7\pm0.3$ \\ % with self-calibration
  2A0335 & 110.35 & 03:38:40.548 & +09:58:12.07 & $6.81\pm0.04$ \\ % with self-cal
       & 335.15 & & & $1.2\pm0.2$ \\ % position is consistent with CO(1-0), too faint for self-cal
  RXJ0821 & 98.80 & 08:21:02.198 & +07:51:48.81 & $0.11\pm0.03$ \\ % Too faint for self-cal.  Additional source at 08:21:00.807, +07:51:27.356 coincides with neighbouring galaxy
       & 304.70 & & & $\sim4$ \\ % Blobby, problems with calibrator
  RXCJ1504 & 101.69 & 15:04:07.518 & -02:48:16.63 & $8.38\pm0.02$ \\ % no self-cal
       & 291.36 & & & $5.33\pm0.06$ \\
  Phoenix & 225.09 & 23:44:43.902 & -42:43:12.53 & $2.5\pm0.1$ \\
  A1664 & 96.27 & 13:03:42.567 & -24:14:42.23 & $2.47\pm0.07$ \\
       & 300.63 & & & $1.2\pm0.1$ \\
  A1835 & 97.89 & 14:01:02.083 & +02:52:42.65 & $1.26\pm0.05$ \\
       & 282.91 & & & $0.7\pm0.2$ \\
  A262 & 235.74 & 01:52:46.456 & +36:09:06.42 & $3.22\pm0.07$ \\
\hline
\label{tab:cont}
\end{tabular}
\end{center}
\end{table}

% Need to note that Phoenix filaments are more massive than this (not including the base of each one, which overlaps with disk)

\section{Discussion}
\label{sec:discussion}

% Filament vs disk-dominated morphologies

The molecular gas reservoirs in the central cluster galaxies form a
continuous distribution in morphology from filament-dominated to
disk-dominated (Fig. \ref{fig:montageflux}).  Filament-dominated
sources, such as the Perseus cluster, A1795 and PKS\,0745, feature
molecular filaments extending a few to tens of kpc, which
encompass the majority of the molecular gas mass.  In contrast,
the molecular gas in disk-dominated sources, Hydra A and A262,
is concentrated in a circumnuclear rotating disk with wispy filaments
comprising $\sim10\%$ of the molecular gas mass.  Disk-dominated systems appear rare and the majority of the ALMA
targets lie in a continuum between these extremes with the molecular
gas split more evenly between a central peak and extended filaments.

% Filaments entrained around cavities, smooth gas flows

The extended molecular filaments are clearly entrained around or drawn
up in the wakes of radio bubbles inflated by the AGN in at least six
of the twelve central galaxies.  Although several molecular structures
in the remaining targets appear aligned with X-ray cavities, any clear
morphological link is weakened by the mismatch in spatial scales
probed by ALMA and \textit{Chandra}.  A1795 represents the most
spectacular example of this entrainment.  Molecular gas in the
central galaxy is exclusively projected around the outer edges of the
radio lobes and particularly bright clumps of cold gas are coincident
with notches and bends in the radio lobes (Fig. \ref{fig:a1795}).
Filaments typically have smooth velocity gradients along
their lengths spanning a few hundred $\kmps$ and narrow FWHM
$<100\kmps$.  The filaments are therefore gas flows tracing
streamlines around and behind the radio bubbles, which may retain the
velocity structure of the rising bubble or decouple and slowly fall back
towards the galaxy centre.

% Low velocities, circulation flows

The molecular gas velocities in these central galaxies are low and
fall significantly below the galaxy's stellar velocity dispersion.
The gas flows are moving too slowly to escape the central galaxy and
even the highest velocity structures at $\pm600\kmps$ in the Phoenix
cluster and A1664 will remain bound.  With the exceptions of the large
circumnuclear gas disks, the molecular gas structures have low
velocities and dispersions and are therefore not settled in the
gravitational potential.  The gas flows are also not in free-fall and
must be decelerated, potentially by some combination of ram pressure or magnetic
fields.  The distribution in morphology from disk- to
filament-dominated sources suggests a slowly varying, dynamic
environment dictated by the episodic activity of the jet-inflated
bubbles (eg. in the Perseus cluster,
\citealt{Lim08,Salome06,Salome11}).

\subsection{Origin of the molecular gas in central cluster galaxies}

% Total molecular gas mass vs. cooling in filaments
% Quote trends for the former?  

\begin{figure}
\centering
\includegraphics[width=0.9\columnwidth]{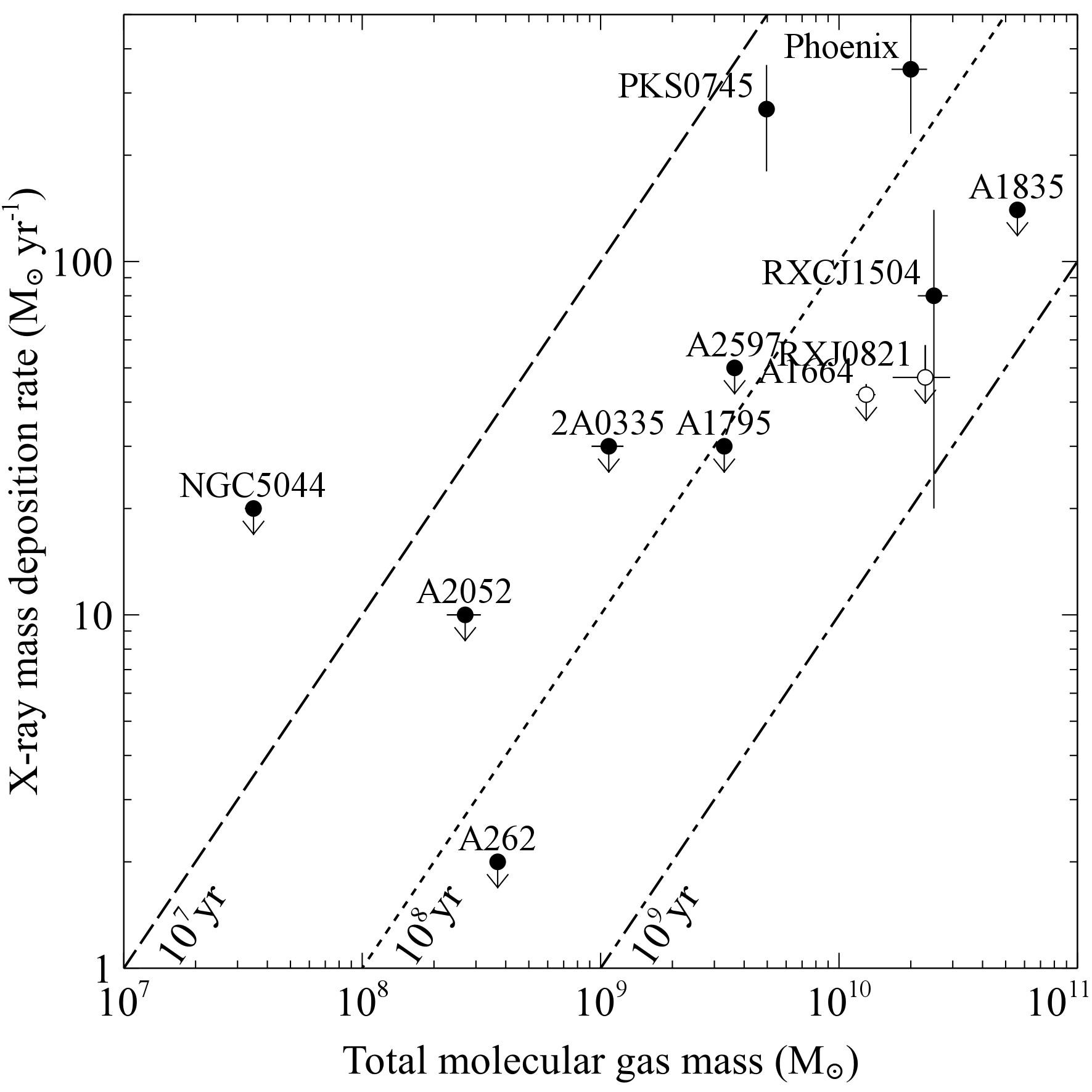}
\caption{Total molecular gas mass vs XMM RGS (solid points) and
  Chandra (open points) upper limits on the mass deposition rate
  cooling from the X-ray hot atmosphere.  The time required for the
  X-ray cooling rate to supply the observed molecular gas mass is
  demonstrated by the dashed ($10^7\yr$), dotted ($10^8\yr$) and
  dash-dotted lines ($10^9\yr$).}
\label{fig:mdotvsCOmass}
\end{figure}

\begin{figure}
\centering
\includegraphics[width=0.9\columnwidth]{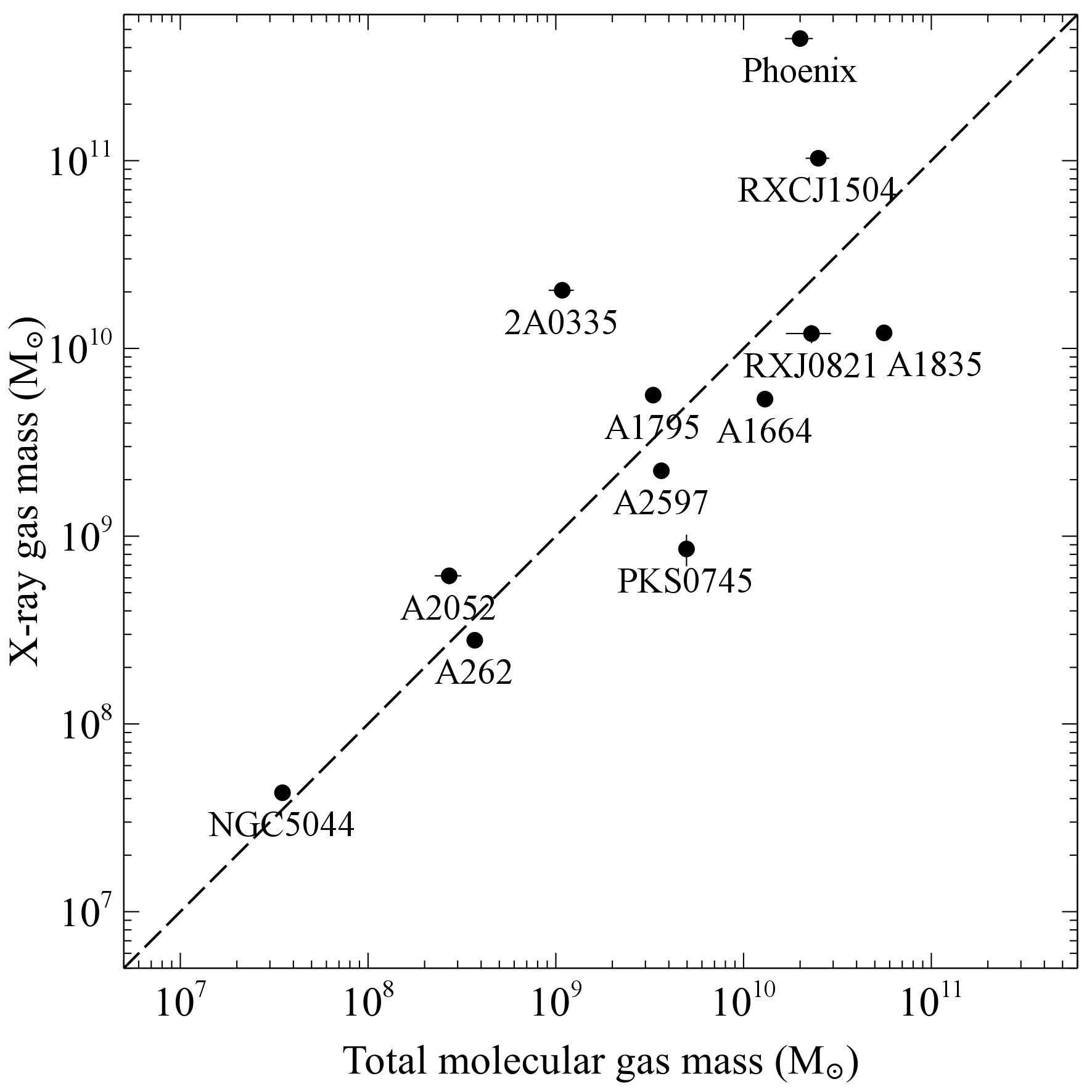}
\caption{Total molecular gas mass vs the X-ray gas mass within the region covered by the molecular emission detected with ALMA (typically $5-15\kpc$).  The molecular gas mass is comparable to the X-ray gas mass in this region as shown by the dashed line.  We note that the apparent inconsistency of the central galaxy in the Phoenix cluster with this trend could indicate that a Milky Way $X_{\mathrm{CO}}$ factor is also appropriate for this source (section \ref{sec:mass}), which would imply upward of $10^{11}\Msun$ of molecular gas.}
\label{fig:xrayvsCOmass}
\end{figure}

When the radiative cooling time of the surrounding hot atmosphere
falls below a Gyr, the central galaxy lights up with star
formation and ionized and molecular line emission from a burgeoning
reservoir of cool gas (\citealt{Rafferty08,Cavagnolo08,Pulido18}).
These clear correlations have been consistently supported by observations
for decades (eg. \citealt{Heckman81,Hu85,Heckman89,Johnstone87} and
for reviews see \citealt{McNamara12,Fabian12}).  ICM cooling is now
included as the primary source of cool gas clouds in models and
simulations of AGN feedback in clusters
(\citealt{PizzolatoSoker05,Gaspari13,Li14,Voit17}).  The substantial
cold gas masses ($\times10^{9-10}\Msun$) and star formation rates
(several to hundreds of solar masses) in the most massive central cluster
galaxies cannot be sustained by stellar mass loss or gas stripped from
donor galaxies.  Not even gas-rich spirals can supply gas at this
level and these sources are rare in the cores of clusters, where
galaxies are predominantly devoid of cold gas and star formation
(eg. \citealt{Best07}).  Although these mechanisms will make some
contribution (eg. \citealt{Sparks89,Voit11}), the strong trends
between the X-ray cooling time and the molecular gas mass and star
formation rate are very difficult to account for without requiring
significant gas cooling from the hot atmosphere.

Fig. \ref{fig:mdotvsCOmass} compares the total molecular gas mass
(see appendix) with the best limits on the cooling rate from
the X-ray hot atmosphere.  The strongest X-ray constraints are
produced from clear detections of Fe\,XVII emission and other key
species in XMM RGS observations, which originate in gas cooling below
$1\keV$
(eg. \citealt{Peterson01,Tamura01,Kaastra01,PetersonFabian06}).  XMM
RGS measurements were not available for A1664 and RXJ\,0821, we
therefore used Chandra limits on the cooling rate within $\sim30\asec$
radius (\citealt{BayerKim02,Calzadilla18}).  Based on the limits on
the X-ray cooling rates, the molecular reservoirs typically form on
timescales of $10^8\yr$ or more.  This timescale may be overestimated
if the in situ cooling rate is increased by non-radiative cooling,
where hot ionizing plasma penetrates the cold gas filaments
(\citealt{Fabian02,Soker04,Fabian11}).  However, a significant
fraction of the molecular gas is likely consumed in star formation and
deeper X-ray observations will place stronger limits on the cooling
rate, which will lengthen the formation timescale.

For molecular filaments that form as gas cools in the wakes of
buoyantly rising radio bubbles, the timescales are much more limited.
The buoyant rise time for the bubbles in our sample is typically
$10-30\Myr$.  The limits on X-ray cooling rates of $30-300\Msunpyr$
are currently high enough to supply the inferred molecular gas masses
of the filaments, typically a few $\times10^{8-9}$ on these
timescales.  However, this would require a large fraction of the hot
X-ray atmosphere within the central galaxy to cool on these
$10-30\Myr$ timescales.  Fig. \ref{fig:xrayvsCOmass} compares the mass
of X-ray gas within the extent of the detected molecular emission
($\sim5-15\kpc$) with the total molecular gas mass (see
\citealt{Pulido18} for the comparison using single dish observations).
The molecular gas mass is within a factor of a few of the X-ray gas
mass within the same region for the majority of our sample.  Cooling
to supply these molecular reservoirs would then deplete the X-ray
atmosphere within this region and require significant inflow.  Such an
inflow would oppose the observed metal-rich hot gas flows along the
jet axis (section \ref{sec:inflowoutflow}).  It is more likely that a lower
level of X-ray cooling occurs over a larger region and this feeds more
extended, fainter molecular filaments that are not yet detected in
early shallow ALMA observations.  This is supported by the greater
extent of the ionized gas filaments, which are closely associated with
the molecular filaments, and the fainter but far more extended
molecular structures detected in IRAM and CARMA observations of the
nearby, bright clusters Perseus and A1795
(\citealt{Salome06,Salome11,McDonald12A1795}).

%Observation suggests that a short central cooling time ($<10^{9}\yr$),
%the ability of bubbles to lift atmospheric gas, and the amount of
%cooling atmospheric gas all influence the level of molecular gas
%observed in these systems.

%%% Needs discussion of cooling to molecular phase here %%%

The formation and structure of cold gas clouds in the intracluster
medium has been considered and modelled in detail by Ferland et
al. (\citeyear{Ferland94,Ferland09}).  Many unknowns remain and a model
that reproduces the low-ionization spectra of the cool gas nebulae in
central cluster galaxies has been a long-standing challenge (reviewed
by eg. \citealt{Johnstone07}).  The formation of molecular hydrogen is
the slowest step and must be catalyzed by dust grain surfaces to occur
on the bubble rise timescales (eg. \citealt{Ferland09}).  Although
dust grains will be sputtered rapidly ($<1\Myr$) in cluster
atmospheres (\citealt{Draine79,Dwek92}), and we would expect cooling
X-ray gas to be dust-free, many of the molecular filaments are
observed to be spatially coincident with dust lanes
(eg. \citealt{Mittal11,Mittal12,Russell16, Russell17, Vantyghem16,Vantyghem18}).  Little
new star formation occurs in the majority of the filaments.  However,
dust could be distributed locally in the ejecta of the central
galaxy's older stellar population (\citealt{Voit11}), form in situ
within cooling gas clouds (\citealt{Fabiandust94}) or have been lifted
from the galaxy centre and shielded in dense gas clumps.

\subsection{Gas flows lifted in the wakes of radio bubbles}
\label{sec:discbubbles}

% Direct uplift unlikely => extended fraction and no strong variations in XCO (discussed in outflows section)
% Stimulated cooling idea => gas must be lifted to cool

Based on the observed close entrainment of the molecular gas flows
around radio bubbles (sections \ref{sec:morphology} and
\ref{sec:vel}), \citet{McNamara16} proposed the stimulated feedback
model where molecular clouds cool from low entropy X-ray gas lifted in
the wakes of buoyant radio bubbles.  Low entropy overdense gas blobs are expected to sink rapidly in a hot atmosphere to a radius where the ambient density is similar, and refind their equilibrium, before they can condense (\citealt{Nulsen86}).  Therefore, for the blobs to become thermally unstable, their radiative cooling time ($t_{\mathrm{cool}}$) must be shorter than the time it takes them to sink to their equilibrium position ($t_{\mathrm{infall}}$).  Low entropy X-ray gas may therefore cool to low temperatures when lifted by radio bubbles to an altitude where $t_{\mathrm{infall}}>t_{\mathrm{cool}}$.
%From \citet{Nulsen86}, the infall time of an overdense, X-ray gas blob can be estimated from its terminal speed, which is determined by balancing drag against the net buoyant force.  
%\begin{equation}
%\rho_{\mathrm{e}} A v_{\mathrm{t}}^2\simeq \left( \rho - \rho_{\mathrm{e}}\right)Vg,
%\end{equation}
%\noindent where $\rho_{\mathrm{e}}$ is the ambient density, $\rho$ is the density of the blob, $g=v_{\mathrm{K}}^2/R$ is the acceleration due to gravity and $V$ and $A$ are the volume and effective cross section of the blob respectively.  The terminal speed $v_{\mathrm{t}}$ is then given by:
%\begin{equation}
%v_{\mathrm{t}} = v_{\mathrm{K}} \sqrt{\frac{\left(\rho - \rho_{\mathrm{e}}\right)r}{\left(\rho_{\mathrm{e}}R\right)}},
%\end{equation}
%\noindent where $r=V/A$ is a measure of the blob's size.  If the terminal speed is smaller than free fall, the ambient gas can slow a blob significantly. 
Since $t_{\mathrm{infall}}\geq t_{\mathrm{ff}}$ the free-fall time, the maximum radius that a radio bubble would need to lift cooler, denser gas to is where $t_{\mathrm{cool}}\simeq t_{\mathrm{ff}}$.   This is typically
a few tens of kpc for our sample.  Although we cannot measure the velocities of the X-ray gas, the observed molecular gas velocities
are considerably lower than the expectations for free fall in these
central cluster galaxies (section \ref{sec:inflowoutflow}).  Therefore, the
infall time is likely significantly longer than the free fall time and
thermal instability will be triggered when low entropy gas is lifted smaller
distances.

%\begin{figure}
%\centering
%\includegraphics[width=0.9\columnwidth]{outflowenergetics.jpg}
%\caption{Energy required to lift the molecular gas in each filament compared with the $4PV$ energy required to inflate the corresponding radio bubble against the pressure of the hot atmosphere.  The fraction of the cavity energy required to lift the molecular gas is demonstrated by the dashed (100\%), dotted (10\%) and dash-dotted lines (1\%).  The bubble volumes, and therefore the bubble energies, are typically uncertain by at least a factor of a few (see eg. \citealt{McNamara12}).}
%\label{fig:outflowenergy}
%\end{figure}

\begin{figure}
\centering
\includegraphics[width=0.9\columnwidth]{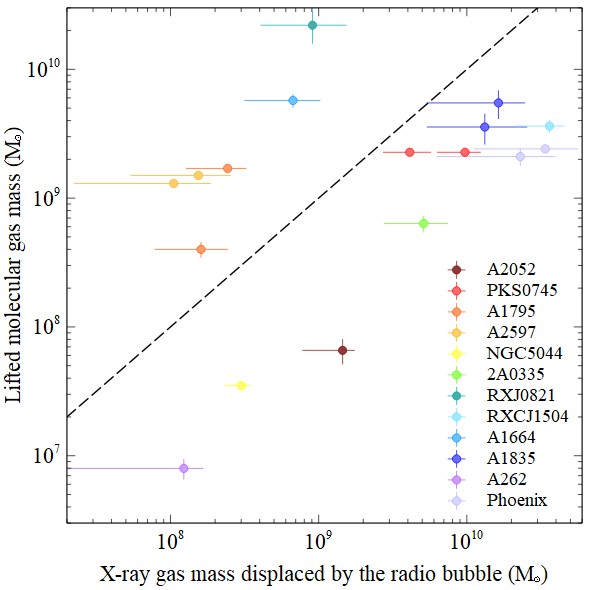}
\caption{Molecular mass of each filament plotted against the X-ray gas mass
  displaced by the corresponding radio bubble.  The dashed line
  denotes equal filament and displaced gas masses.  Note that the uncertainty of at least a factor of a few on the bubble
  volumes propagates into similar
  uncertainties in the displaced gas mass.}
\label{fig:outflowmass}
\end{figure}

Stimulated feedback naturally explains the morphology and velocity
structure of the molecular filaments, the close coupling with the
radio bubbles and the large fraction of the molecular gas lying in
extended filaments.  
Radio bubbles lift material in their wakes through buoyancy and, by Archimedes' principle, cannot lift more gas than they
displace.  The displaced mass can be determined from the size of the
cavities in the X-ray hot atmosphere, assuming spherical or prolate
ellipsoids, and the density of the ambient hot gas from spectral
fitting (eg. \citealt{Cavagnolo09}).  Fig. \ref{fig:outflowmass}
compares the molecular gas mass in each filament that must be lifted by the bubbles with the mass of hot
gas displaced by the corresponding radio bubble.  %The total mass of gas that must be lifted by the bubbles is
%given by the sum of the gas mass in the molecular filament and the
%metal-rich hot outflow (section \ref{sec:outflows}).  Therefore, Fig. \ref{fig:outflowmass} also shows the total mass of gas that must be lifted, which is the molecular gas mass plus the hot outflow gas mass for each filament.  

Both the cold and displaced gas masses have uncertainties of at least a factor of a few.  For the cold gas mass, the uncertainty stems from the $X_{\mathrm{CO}}$ factor (section \ref{sec:mass}) but also from difficulties in separating each filament from other structures.   The filaments may also subsequently be growing in mass due to interpenetration of the hot and cold gas (\citealt{Fabian11Fil,Liu19}).  For the displaced gas mass, the uncertainty in the bubble volume dominates.  The displaced gas mass could also be significantly underestimated if some cavities have collapsed or the filaments were formed by multiple generations of bubbles.  Deep Chandra X-ray observations of nearby clusters can reveal a series
of X-ray cavities at larger radii, where the outer cavities were
generated by previous AGN outbursts and have buoyantly risen through
the hot atmosphere.  Whilst the radio emission from the bubble's relativistic contents has spectrally aged and can only be detected in lower frequency observations, the bubble is still visible as a surface brightness depression or cavity in X-ray observations.  Extended molecular filaments are detected toward
multiple generations of X-ray cavities in the nearby Perseus cluster, which can be studied in detail.  Extended luminous ionized gas filaments, which are spatially
correlated with the molecular gas, are also associated with outer cavities
in other nearby systems (eg. \citealt{Salome11}).  Similarly, in A1795, a large outer bubble is detected in X-rays and low frequency radio observations (\citealt{Crawford05,Kokotanekov18}).  This outer bubble has displaced more than an order of magnitude more hot gas than the two inner bubbles in this system.

Given the uncertainties, we conclude that the displaced and cold gas masses are of roughly comparable magnitude.  Larger radio bubbles are also typically associated with more massive cold gas filaments.  This is consistent with the stimulated feedback model.  Although the discrepancies can generally be attributed to the expected scatter, there are known exceptions.  For example, MS\,0735 hosts a particularly powerful radio bubble outburst in excess of $10^{46}\ergps$ that has displaced $>10^{12}\Msun$ of hot gas but the molecular gas supply is less than $3\times10^9\Msun$ (\citealt{Salome08MS07,McNamara09,Vantyghem14}). The link between the radio bubble activity and molecular filament formation may therefore be more complex.

\subsection{AGN fuelling}

\begin{figure}
  \centering
  \includegraphics[width=0.9\columnwidth]{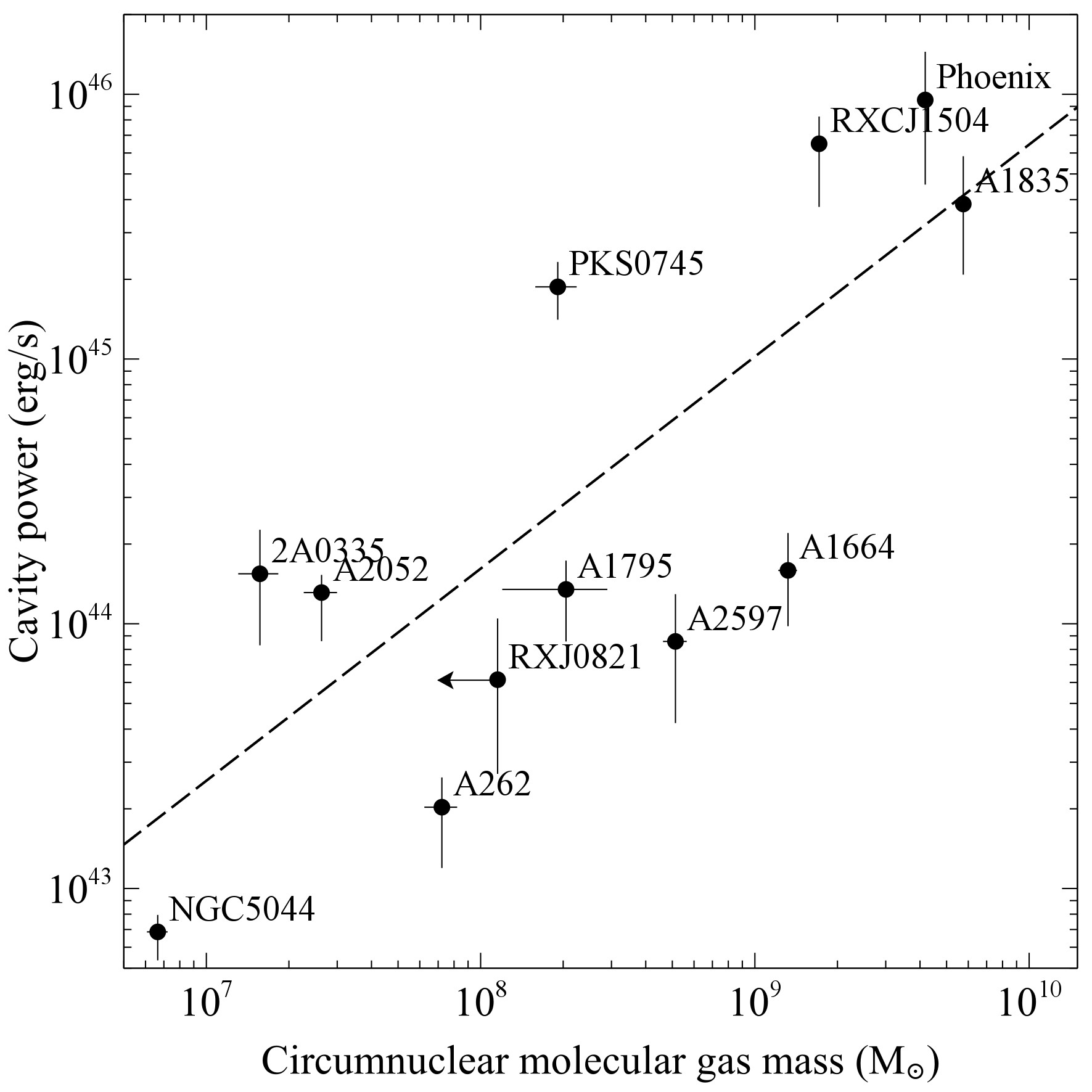}
  \caption{Molecular gas mass in a single ALMA synthesized beam centred on the AGN compared with the jet power required to inflate the innermost radio bubbles (X-ray cavities).  The BCES (Y$\vert$X) fit to the data points is shown by the dashed line.}
  \label{fig:fuelling}
\end{figure}

The observed balance between the AGN heating and radiative cooling
rates in galaxy cluster atmospheres must be mediated through fuelling
of the central SMBH.  The accretion rate must be sensitive to
overcooling or overheating on larger scales so that the AGN activity
compensates on appropriately short timescales.  Although the AGN must
accrete at some level from the X-ray atmosphere (Bondi accretion,
eg. \citealt{Allen06,Russell13}), the hot gas supply is not sufficient to
power the most energetic radio jet outbursts
(eg. \citealt{Rafferty06,McNamara11}).  Molecular gas provides an
alternative and plentiful supply of fuel and ALMA can now begin to
resolve the circumnuclear structures from more extended filaments in
these systems.

Although the ALMA observations of this sample were optimized for
extended structure on kpc-scales, this still represents a significant
improvement in spatial resolution over previous studies utilizing
single dish observations (eg. \citealt{Pulido18}).
Fig. \ref{fig:fuelling} compares the molecular gas mass within a
single synthesized beam centred on the AGN (see appendix) with
the jet power from the innermost radio bubbles.  Using the BCES
(Y$\vert$X) estimator from \citet{Akritas96}, we determine the
best-fit powerlaw model

\begin{equation}
\mathrm{log}\left(P_{\mathrm{cav}}\right) = \left( 0.80\pm0.16\right)\mathrm{log}\left(M_{\mathrm{H_{2},nuc}}\right) + \left( 37.8\pm1.4\right),
\end{equation}

\noindent where $P_{\mathrm{cav}}$ is the X-ray cavity power and
$M_{\mathrm{H_{2},nuc}}$ is the circumnuclear molecular gas mass.  We
also calculate the Spearman rank correlation coefficient of 0.75 with
p-value 0.007, which suggests a tentative correlation.  However, this
does not account for the large uncertainties in the X-ray cavity
power.  If we employ a bootstrapping method to resample the data and
perturb the resampled values according to the uncertainties, we do not
find a significant correlation between the circumnuclear molecular gas
mass and the AGN jet power as measured by the X-ray cavities.
Although this may be partly due to the limited number of systems
observed so far with ALMA (particularly lower mass systems) and large uncertainties on the X-ray cavity
power, higher spatial resolution ALMA measurements
will be required to probe the circumnuclear structure.  

Smaller scale structure may also be probed by absorption lines detected against the nuclear continuum emission.   CO absorption lines have been detected in three sources considered here, NGC\,5044, A2597 and Hydra A.  These narrow features are consistent with obscuring giant molecular clouds along the line of sight, which are likely located within a kpc of the nucleus (\citealt{David14,Tremblay16,Rose19}).

% X-ray gas not a sufficient source of fuel in most powerful systems
% Single dish obs. probed large scales

% Large range of spatial scale probed by single beam

\section{Conclusions}

Central cluster galaxies with short atmospheric cooling times are rich
in molecular gas, with masses from $10^9\Msun$ to nearly
$10^{11}\Msun$.  Unlike gas-rich spiral galaxies, molecular gas in
these massive galaxies is rarely found in ordered motion, such as a
disk or ring.  Little gas is therefore centrifugally supported.
Instead, the molecular gas is filamentary and/or in turbulent-like
motion in the host galaxy.  Their morphologies and velocity fields
give strong clues to the origins of molecular clouds and their
relationship to radio-mechanical feedback in galaxies.

Molecular gas filaments are found preferentially around or beneath
rising X-ray bubbles formed by radio jets.  Their ensemble velocity
dispersions lie far below the host galaxy's stellar velocity dispersions.
Likewise, molecular filament bulk velocities lie far below free-fall
speeds.  Their velocity widths are only tens of kilometers per second.
Therefore, the molecular clouds likely formed recently and have had
little time to respond to gravity, and/or they are supported by a
combination of ram pressure and magnetic fields.

Whether molecular clouds are falling in or flowing out is unclear in
any given system as their locations along the line of sight with
respect to the central galaxy are uncertain or unknown.  In some
instances the clouds are clearly moving out (A1835) while in others
the gas may be falling inward (Phoenix).  In all instances, molecular
cloud velocities lie far below the central galaxy's escape speed $\sim
1000\kmps$.  Thus, molecular outflows will eventually stall, return
and circulate within the galaxy.  The association of molecular clouds
and filaments with X-ray cavities and radio lobes indicates that the
molecular flows are being driven by the expanding and rising radio
bubbles.

Two lines of evidence indicate that the molecular clouds originated in
cooling from hot atmospheres.  Firstly, molecular clouds are found
preferentially in central galaxies where the cooling time of the hot
atmosphere lies below $\sim 10^9\yr$.  In addition, in the systems
studied here and elsewhere (eg. \citealt{Pulido18}), the molecular gas
mass correlates with the hot, atmospheric mass within the volume where
molecular clouds are found.  The mass of molecular clouds found in
most systems is also comparable, generally within factors of a few, to
the atmospheric mass displaced and/or lifted outward by the rising
radio bubbles.  This is consistent with the conjecture that molecular
clouds form in the cooling updrafts of rising radio bubbles
(\citealt{Salome11, McNamara14, McNamara16}).  Molecular clouds may
form prodigiously when cooling parcels of gas are lifted to an
altitude where the ratio of their infall time to cooling time falls
below unity, i.e., $t_{\mathrm{cool}}/t_{\mathrm{infall}} \lesssim 1$.

We identify a tentative trend between the unresolved molecular gas
mass surrounding the central AGN and jet power.  However, the
correlation is marginally significant owing to large measurement
uncertainties and small sample size. Higher spatial resolution ALMA
observations are required to probe the circumnuclear structure and
determine more effectively if the AGN activity is fuelled by the
plentiful supply of molecular gas.

\section*{Acknowledgements}
We thank the reviewer for helpful comments that improved the paper.  HRR acknowledges support from an STFC Ernest Rutherford Fellowship and an Anne McLaren Fellowship.  BRM acknowledges support from the Natural Sciences and Engineering Council of Canada and the Canadian Space Agency Space Science Enhancement Program.  ACF acknowledges support from ERC Advanced Grant Feedback 340442.  ACE acknowledges support from STFC grant ST/P000541/1.  ALMA is a partnership of ESO
(representing its member states), NSF (USA) and NINS (Japan), together
with NRC (Canada), NSC and ASIAA (Taiwan), and KASI (Republic of
Korea), in cooperation with the Republic of Chile. The Joint ALMA
Observatory is operated by ESO, AUI/NRAO and NAOJ.  The scientific
results reported in this article are also based on data obtained from the
Chandra Data Archive.  

\section*{Appendix: Individual ALMA targets}

A subset of the figures are reproduced here.  Figures of A1835, NGC5044, RXJ0821, 2A0335 and RXCJ1504, together with the table of best-fit parameters, are available in supplementary material online.

%%% Abell 2052:

\begin{figure*}
\begin{minipage}{\textwidth}
\centering
\includegraphics[width=0.32\columnwidth]{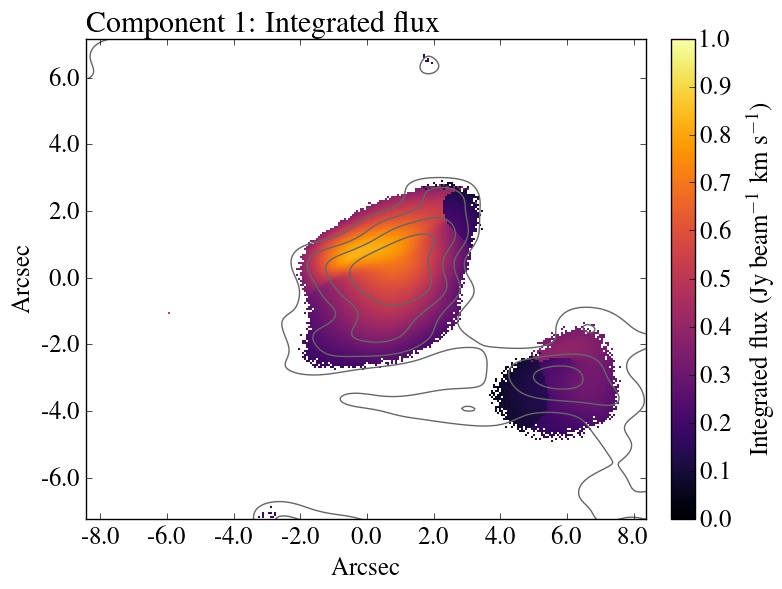}
\includegraphics[width=0.33\columnwidth]{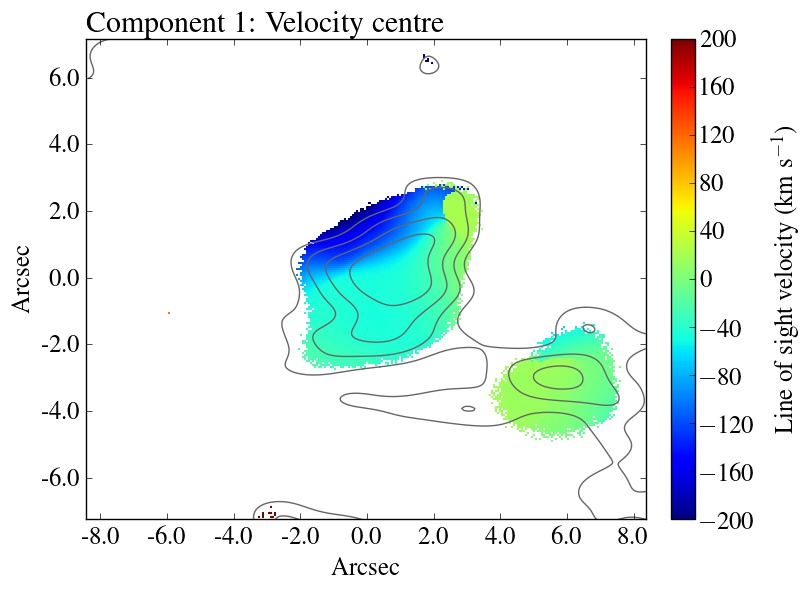}
\includegraphics[width=0.32\columnwidth]{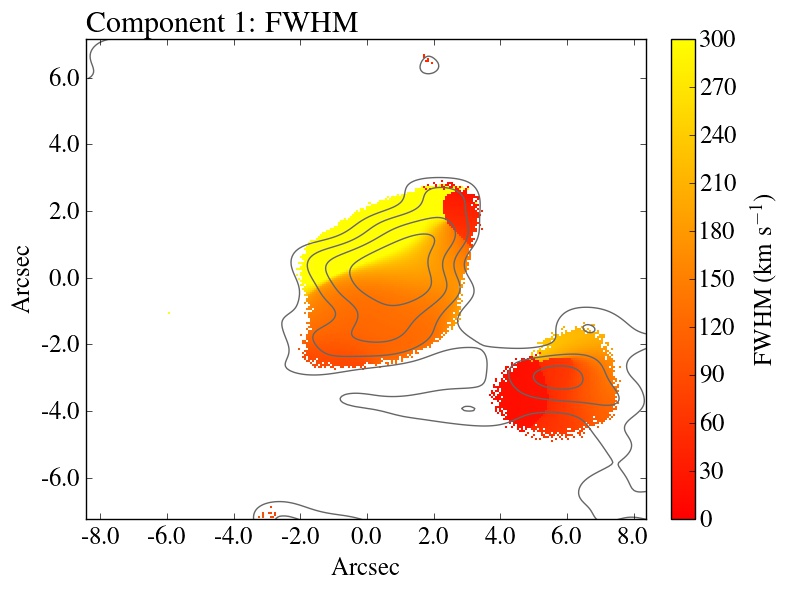}
\caption{Maps of the best-fit integrated intensity (left), velocity centre (centre) and FWHM (right) in A2052 for a Gaussian component detected at $>3\sigma$.  Integrated intensity contours from Fig. \ref{fig:a2052} (right) are overlaid.  The velocity structure of the molecular gas is well-matched to that of the ionized gas on the scales resolved by ALMA (\citealt{Balmaverde18}).  The ionized gas filaments extend around the N radio bubble and the smooth velocity gradients along their lengths can be reproduced with a model for an expanding bubble.}%  The measured velocities and FWHM for the apparent features at the edge of the image have large uncertainties and therefore do not provide further insight on their nature or contradict our conclusion that these are noise. }
\label{fig:a2052maps}
\end{minipage}
\end{figure*}

%%% PKS0745:

% Velocity maps don't look great with radio contours => leave off and note radio in text.
\begin{figure*}
\begin{minipage}{\textwidth}
\centering
\includegraphics[width=0.32\columnwidth]{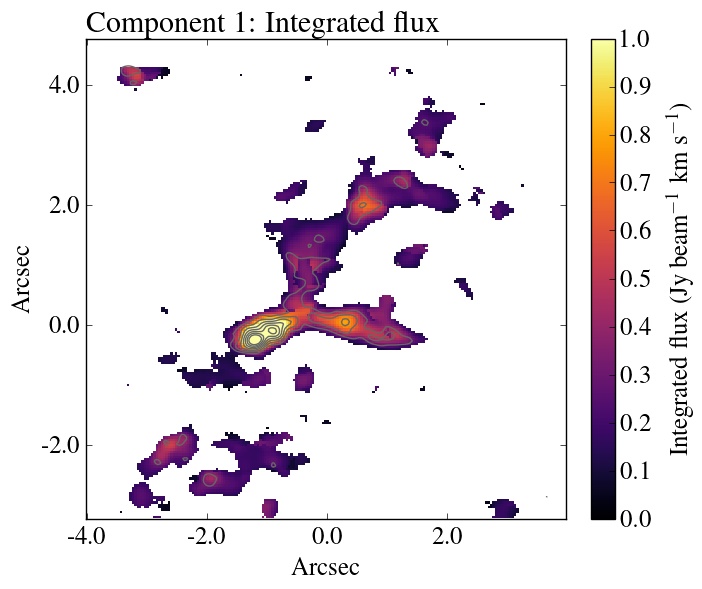}
\includegraphics[width=0.33\columnwidth]{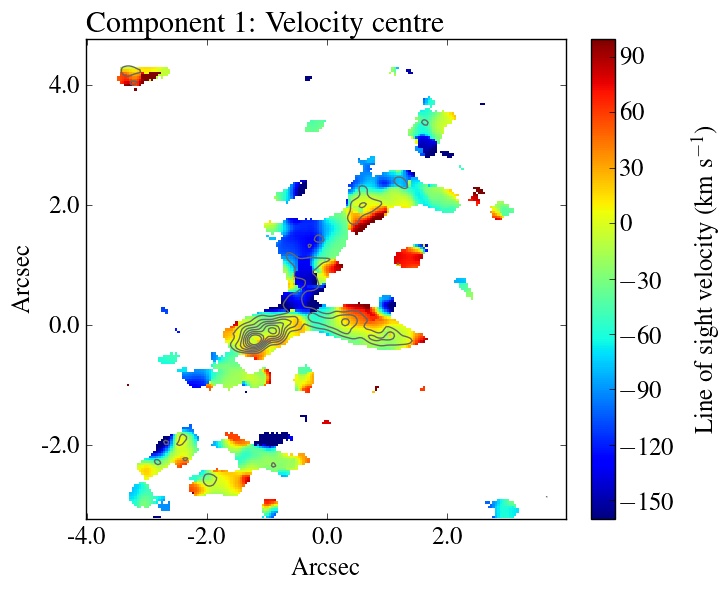}
\includegraphics[width=0.32\columnwidth]{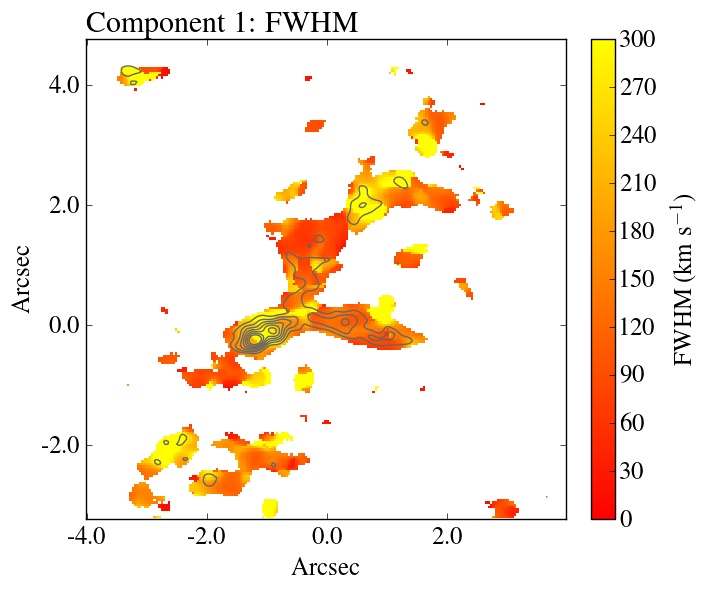}
\caption{Maps of the best-fit integrated intensity (left), velocity centre (centre) and FWHM (right) in PKS0745-191 for a Gaussian component detected at $>3\sigma$.  Integrated intensity contours from Fig. \ref{fig:pks07} are overlaid.  An additional fainter velocity component (not shown) is detected at $0\kmps$ and $-50\kmps$ in synthesized beam-sized regions at the emission peaks of the N and SE filaments, respectively (\citealt{Russell16}).}
\label{fig:pks07maps}
\end{minipage}
\end{figure*}

%%% Abell 1795:


%%% Abell 2597:

\begin{figure*}
\begin{minipage}{\textwidth}
  \centering
\raisebox{0.3cm}{\includegraphics[width=0.29\columnwidth]{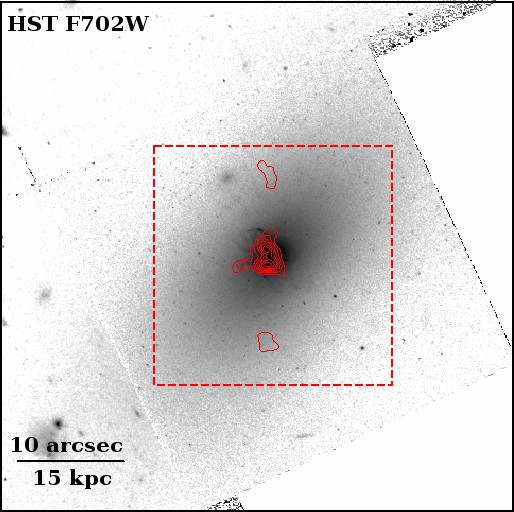}}
\raisebox{0.3cm}{\includegraphics[width=0.29\columnwidth]{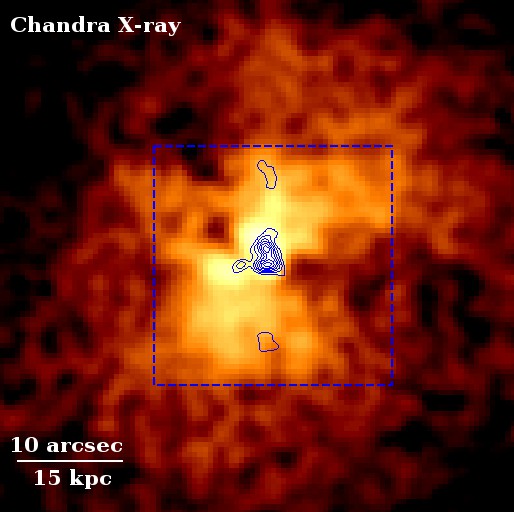}}
\includegraphics[width=0.4\columnwidth]{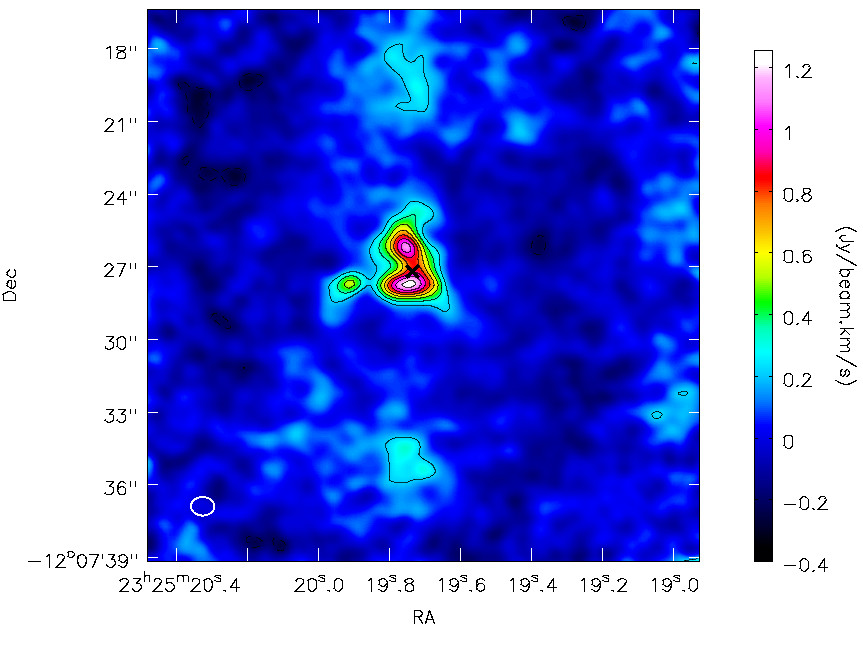}
\caption{Abell 2597.  Left: HST F702W archival image of the central galaxy.  Centre: Chandra X-ray image showing the hot cluster atmosphere.  Right: CO(2-1) integrated intensity map for velocities $-250$ to $+400\kmps$ with contours at $-3\sigma,3\sigma,5\sigma,7\sigma ...$, where $\sigma=0.07\Jypbmkmps$.}
\label{fig:a2597}
\end{minipage}
\end{figure*}
\begin{figure*}
\begin{minipage}{\textwidth}
\centering
\includegraphics[width=0.32\columnwidth]{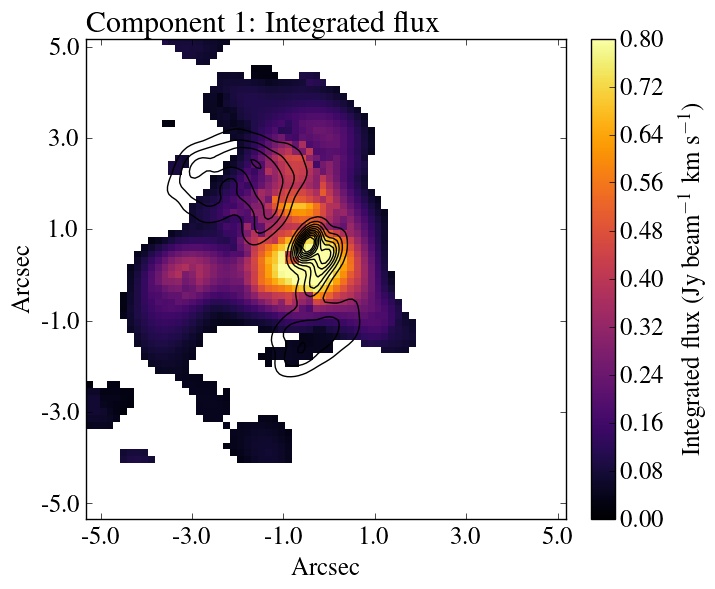}
\includegraphics[width=0.33\columnwidth]{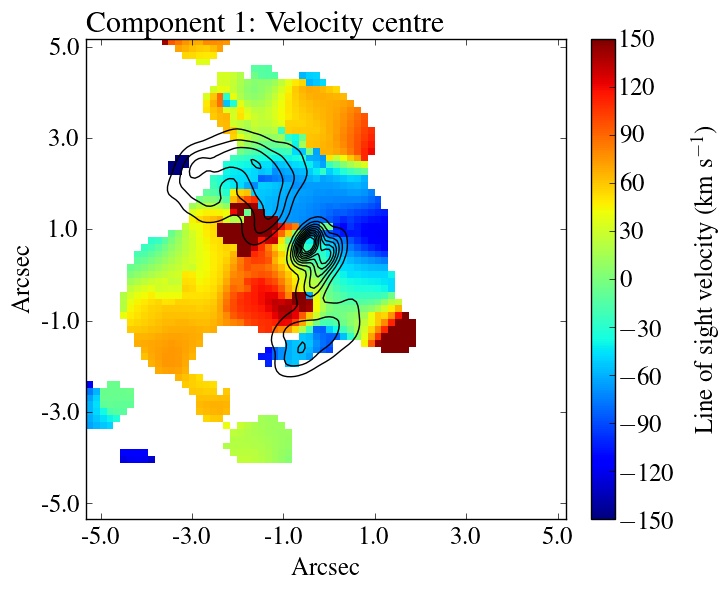}
\includegraphics[width=0.32\columnwidth]{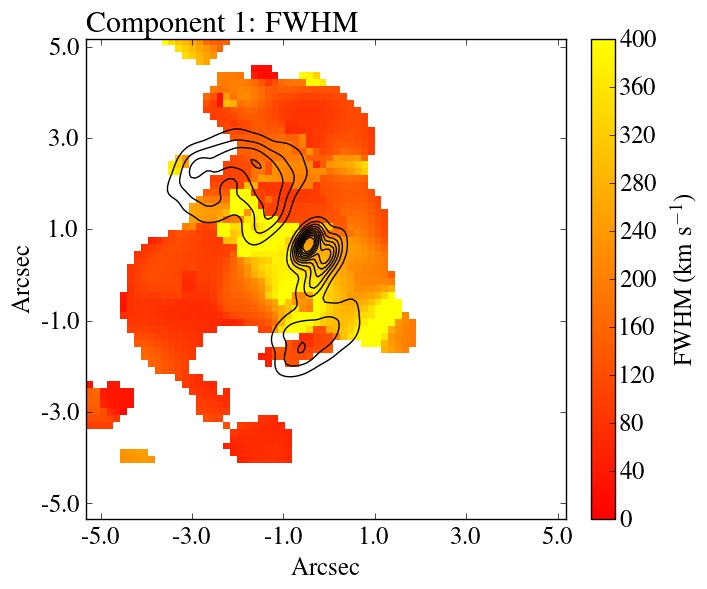}
\includegraphics[width=0.32\columnwidth]{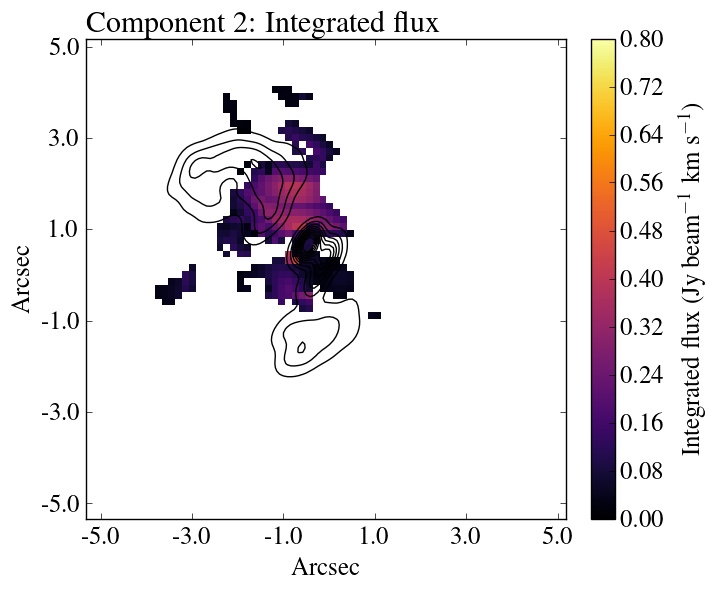}
\includegraphics[width=0.33\columnwidth]{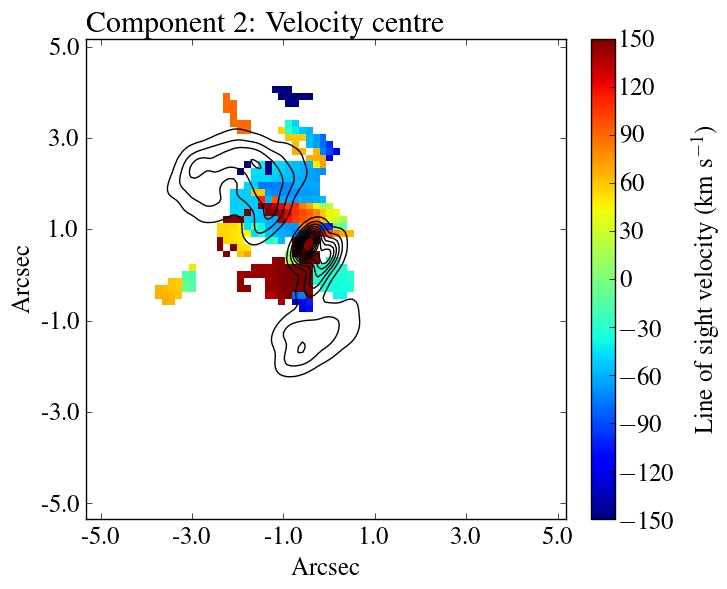}
\includegraphics[width=0.32\columnwidth]{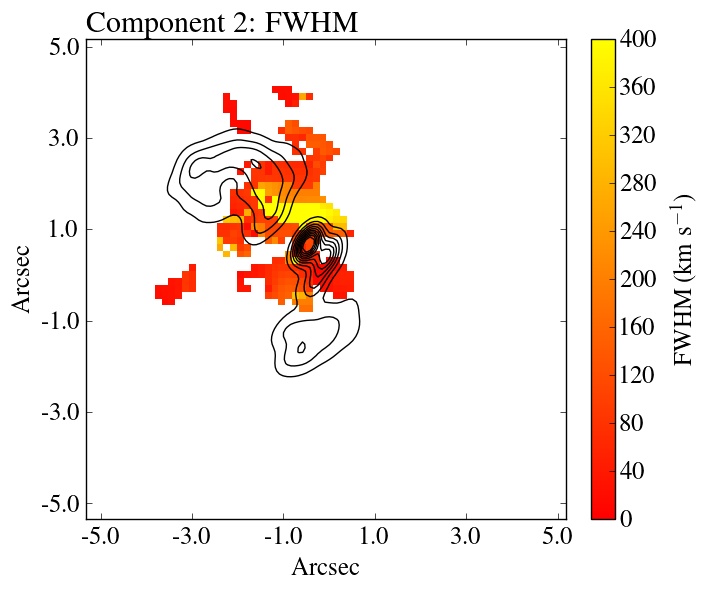}
\caption{Maps of the best-fit integrated intensity (left), velocity centre (centre) and FWHM (right) in A2597 for Gaussian components detected at $>3\sigma$.  VLA $4.9\GHz$ contours are superimposed (\citealt{Sarazin95,Clarke05}).}
\label{fig:a2597maps}
\end{minipage}
\end{figure*}


%%% A1664:

\begin{figure*}
\begin{minipage}{\textwidth}
  \centering
\includegraphics[width=0.2\columnwidth]{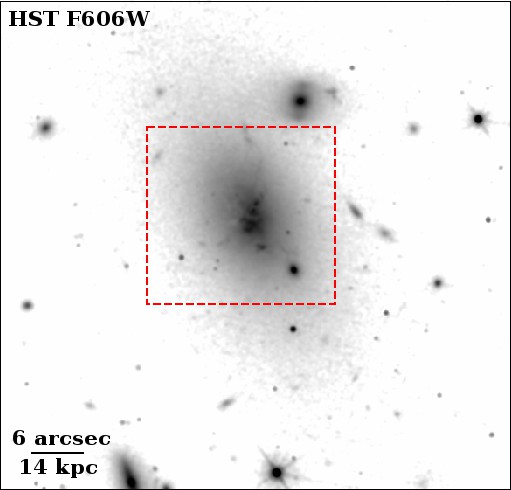}
\includegraphics[width=0.2\columnwidth]{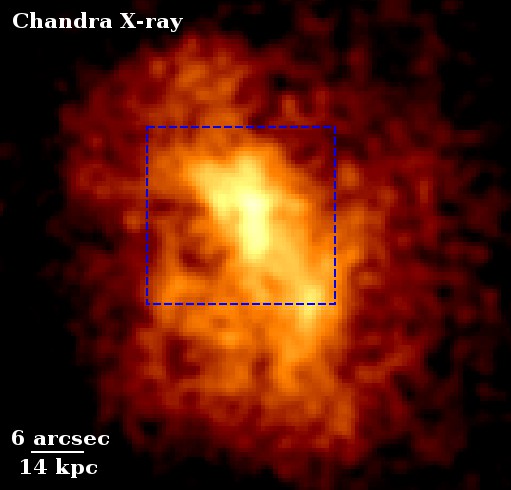}
\raisebox{-0.4cm}{\includegraphics[width=0.29\columnwidth]{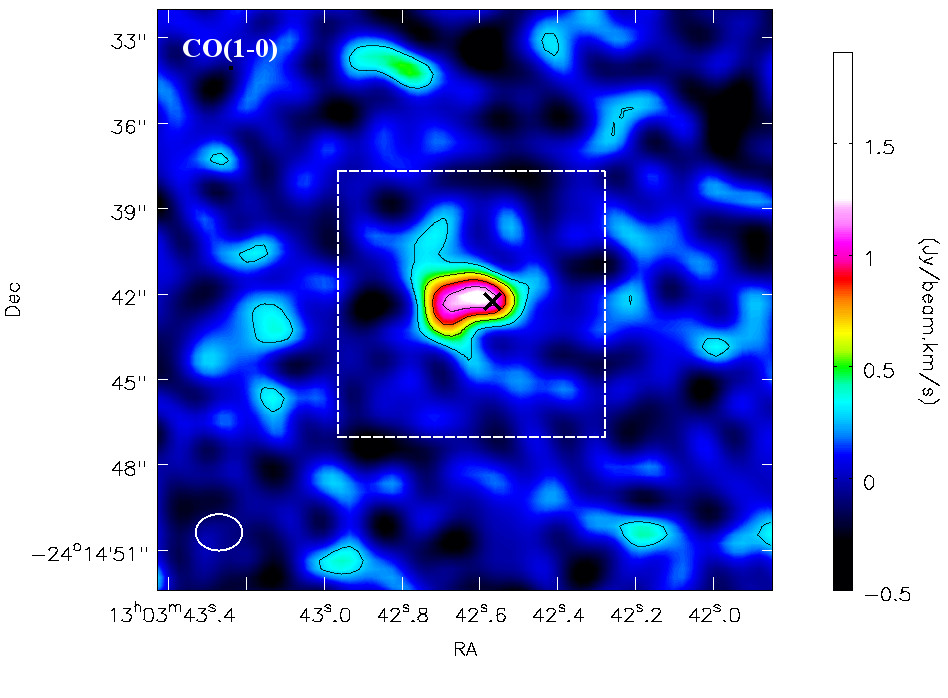}}
\raisebox{-0.4cm}{\includegraphics[width=0.28\columnwidth]{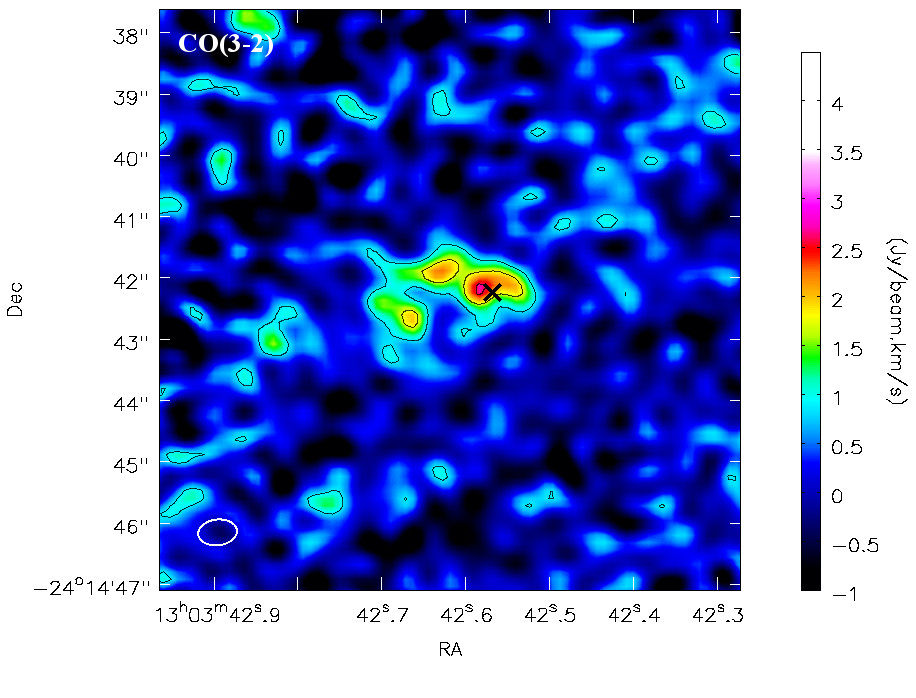}}
\caption{A1664.  Left: HST F606W archival image of the central galaxy.  Centre left: Chandra X-ray image showing the hot cluster atmosphere.  Centre right: CO(1-0) integrated intensity map for velocities $-680$ to $+280\kmps$ with contours at $-2\sigma,2\sigma,4\sigma,6\sigma ...$, where $\sigma=0.14\Jypbmkmps$.  The CO(1-0) field of view is shown as the red and blue boxes in the HST and Chandra images.  Right: CO(3-2) integrated intensity map for velocities $-660$ to $+270\kmps$ with contours at $-2\sigma,2\sigma,4\sigma,6\sigma ...$, where $\sigma=0.43\Jypbmkmps$.  The field of view of the CO(3-2) image is shown in as a white dashed box in the CO(1-0) image.}
\label{fig:A1664}
\end{minipage}
\end{figure*}
\begin{figure*}
\begin{minipage}{\textwidth}
\centering
\includegraphics[width=0.32\columnwidth]{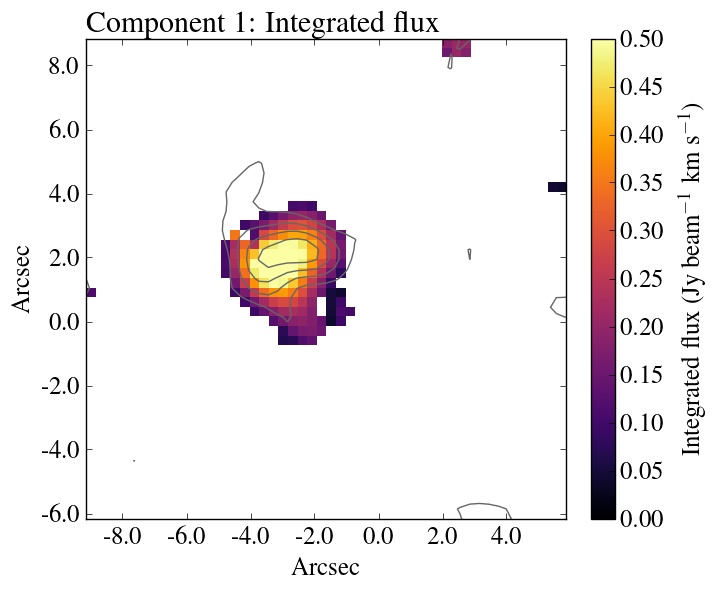}
\includegraphics[width=0.33\columnwidth]{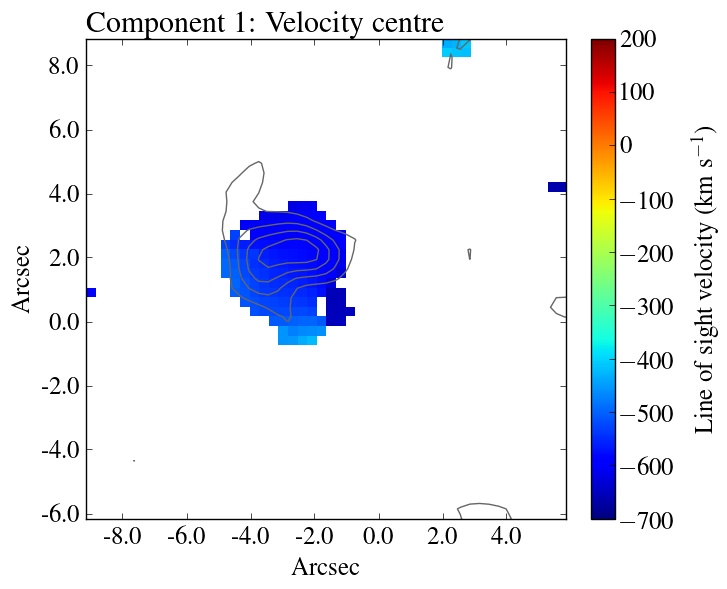}
\includegraphics[width=0.32\columnwidth]{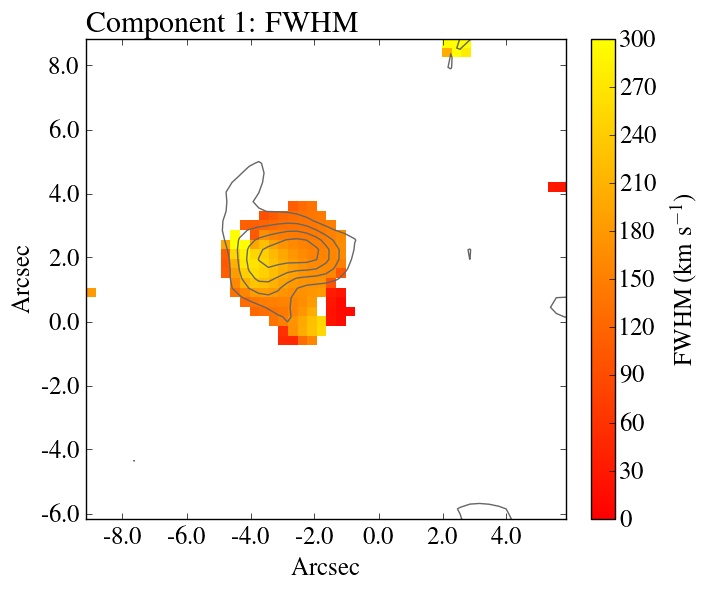}
\includegraphics[width=0.32\columnwidth]{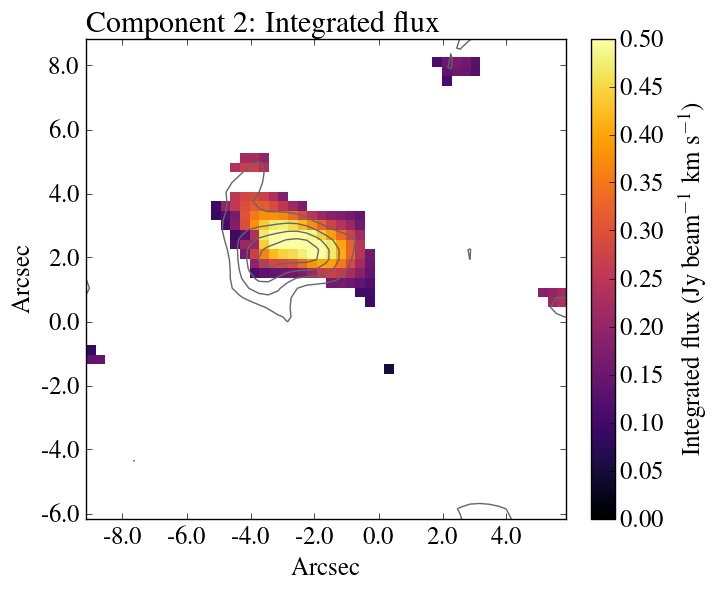}
\includegraphics[width=0.33\columnwidth]{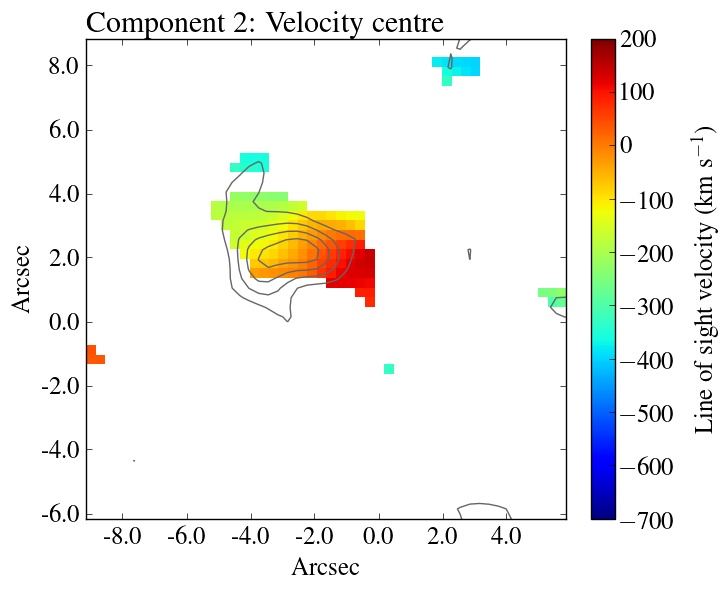}
\includegraphics[width=0.32\columnwidth]{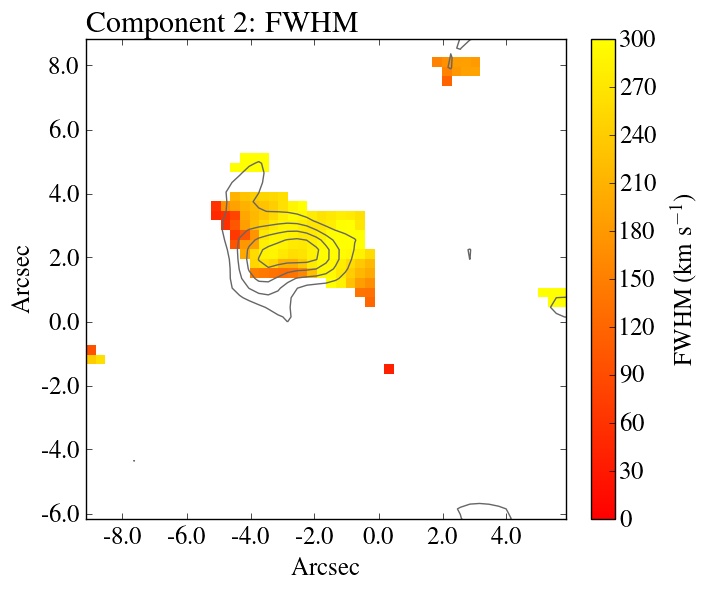}
\caption{A1664 CO(1-0).  Maps of the best-fit integrated intensity (left), velocity centre (centre) and FWHM (right) for Gaussian components detected at $>3\sigma$.  Integrated intensity contours from Fig. \ref{fig:A1664} (right) are overlaid.}
\label{fig:A1664maps}
\end{minipage}
\end{figure*}

\begin{figure*}
\begin{minipage}{\textwidth}
  \centering
\includegraphics[width=0.27\columnwidth]{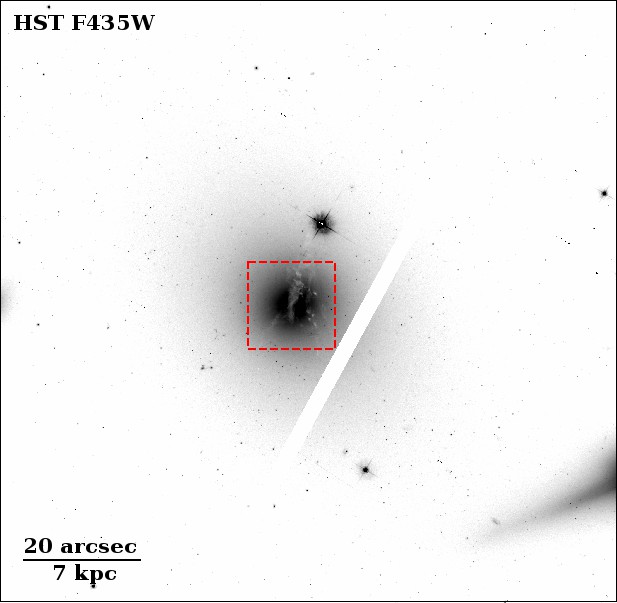}
\includegraphics[width=0.27\columnwidth]{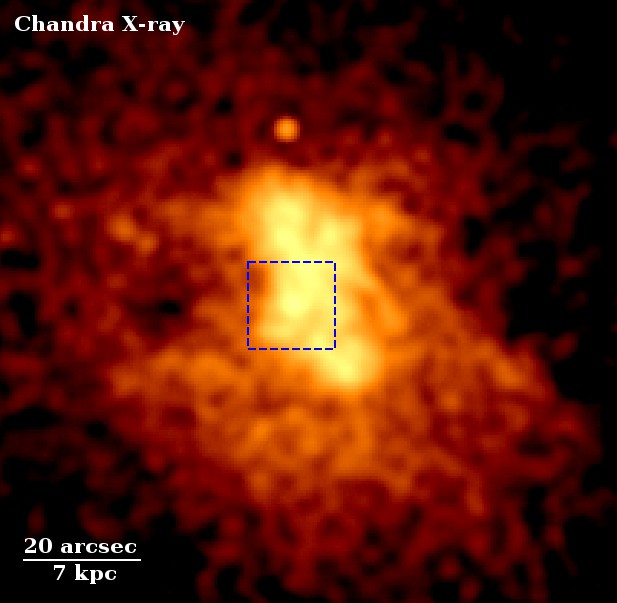}
\raisebox{-0.7cm}{\includegraphics[width=0.39\columnwidth]{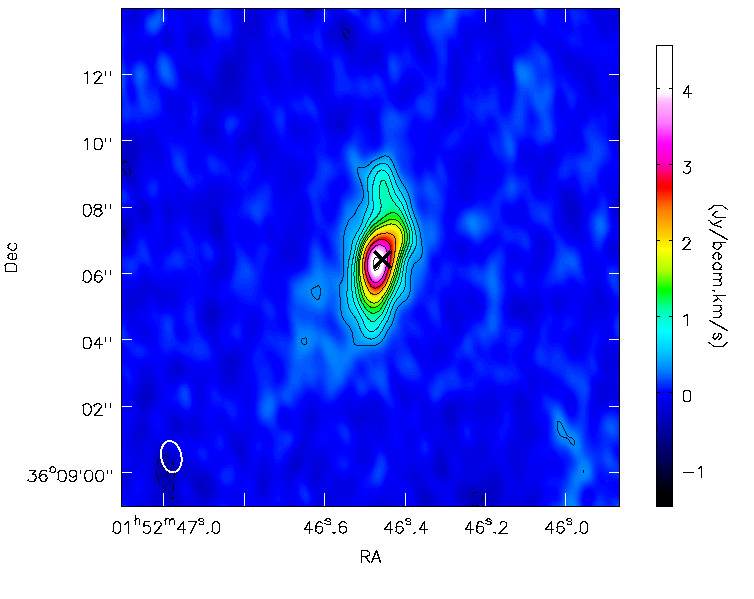}}
\caption{A262.  Left: HST F435W archival image of the central galaxy.  Centre: Chandra X-ray image showing the hot cluster atmosphere.  Right: CO(2-1) integrated intensity map for velocities $-225$ to $+265\kmps$ with contours at $-3\sigma,3\sigma,5\sigma,7\sigma ...$, where $\sigma=0.1\Jypbmkmps$.  The CO(2-1) field of view is shown by the red and blue boxes in the HST and Chandra images.}
\label{fig:A262}
\end{minipage}
\end{figure*}

\bibliographystyle{mnras_mwilliams} 
\bibliography{refs.bib}

\end{document}

%% file: defn_apj.tex
\newcommand{\apj}{ApJ}
\newcommand{\mnras}{MNRAS}
\newcommand{\nat}{Nat}

\newcommand{\araa}{ARA\&A}                % "Ann. Rev. Astron. Astrophys."
\newcommand{\aap}{A\&A}                   % "Astron. Astrophys."
\newcommand{\aj}{AJ}                      % "Astron. J."
\newcommand{\apjs}{ApJS}                  % "Astrophys. J. Suppl. Ser."
\newcommand{\apjl}{ApJ}                   % letter at ApJ

\newcommand{\physrep}{Phys. Rep.}

\newcommand{\nar}{NewA Rev.}

\newcommand{\Mpc}{\rm\; Mpc}
\newcommand{\kpc}{\rm\; kpc}

\newcommand{\km}{\rm\; km}
\newcommand{\m}{\rm\; m}
\newcommand{\cm}{\rm\; cm}

\newcommand{\yr}{\rm\; yr}

\newcommand{\Myr}{\rm\; Myr}
\newcommand{\s}{\rm\; s}

\newcommand{\GHz}{\rm\; GHz}

\newcommand{\K}{\rm\; K}
\newcommand{\Msun}{\hbox{$\rm\thinspace M_{\odot}$}}

\newcommand{\Msunpyr}{\hbox{$\Msun\yr^{-1}\,$}}

\newcommand{\keV}{\rm\; keV}

\newcommand{\erg}{\rm\; erg}
\newcommand{\Jy}{\rm\; Jy}
\newcommand{\mJy}{\rm\; mJy}

\newcommand{\ergps}{\hbox{$\erg\s^{-1}\,$}}

\newcommand{\Jykmps}{\hbox{$\Jy\km\s^{-1}\,$}}
\newcommand{\Jypbmkmps}{\hbox{$\Jy{\rm\thinspace beam^{-1}}{\rm km}\s^{-1}\,$}}
\newcommand{\mJypbm}{\hbox{$\mJy{\rm\thinspace beam^{-1}}$}}
\newcommand{\Kkmps}{\hbox{$\K\km\s^{-1}\,$}}
\newcommand{\kmps}{\hbox{$\km\s^{-1}\,$}}
\newcommand{\kmpspMpc}{\hbox{$\kmps\Mpc^{-1}\,$}}

\newcommand{\keVcmsq}{\hbox{$\keV\cm^{2}\,$}}
\newcommand{\amin}{\rm\; arcmin}

\newcommand{\asec}{\rm\; arcsec}

\newcommand{\psqcm}{\hbox{$\cm^{-2}\,$}}

\newcommand{\pcmcu}{\hbox{$\cm^{-3}\,$}}

\newcommand{\COtoH}{\hbox{$\psqcm(\K\kmps)^{-1}$}}